\documentclass[twocolumn]{aastex62}

\usepackage{amsmath}
\usepackage{multirow}
\received{8$^{\rm{th}}$ September 2020}
\revised{25$^{\rm{th}}$ October 2020}
\accepted{30$^{\rm{th}}$ October 2020}

\begin{document}

\title{Hubble WFC3 Spectroscopy of the Habitable-zone Super-Earth LHS\,1140\,b}

\correspondingauthor{Billy Edwards}
\email{billy.edwards.16@ucl.ac.uk}

\author[0000-0002-5494-3237]{Billy Edwards}
\altaffiliation{These authors contributed equally to this work.}
\affil{Department of Physics and Astronomy, University College London, London, United Kingdom}

\author[0000-0001-6516-4493]{Quentin Changeat}
\altaffiliation{These authors contributed equally to this work.}
\affil{Department of Physics and Astronomy, University College London, London, United Kingdom}

\author[0000-0003-1368-6593]{Mayuko Mori}
\affil{Department of Astronomy, University of Tokyo, Tokyo, Japan}

\author[0000-0002-7771-6432]{Lara O. Anisman}
\affil{Department of Physics and Astronomy, University College London, London, United Kingdom}

\author[0000-0001-8587-2112]{Mario Morvan}
\affil{Department of Physics and Astronomy, University College London, London, United Kingdom}

\author[0000-0002-9616-1524]{Kai Hou Yip}
\affil{Department of Physics and Astronomy, University College London, London, United Kingdom}

\author[0000-0003-3840-1793]{Angelos Tsiaras}
\affil{Department of Physics and Astronomy, University College London, London, United Kingdom}

\author[0000-0003-2241-5330]{Ahmed Al-Refaie} 
\affil{Department of Physics and Astronomy, University College London, London, United Kingdom}

\author[0000-0002-4205-5267]{Ingo Waldmann}
\affil{Department of Physics and Astronomy, University College London, London, United Kingdom}

\author[0000-0001-6058-6654]{Giovanna Tinetti}
\affil{Department of Physics and Astronomy, University College London, London, United Kingdom}

\begin{abstract}

Atmospheric characterisation of temperate, rocky planets is the holy grail of exoplanet studies. These worlds are at the limits of our capabilities with current instrumentation in transmission spectroscopy and challenge our state-of-the-art statistical techniques. Here we present the transmission spectrum of the temperate Super-Earth LHS\,1140b using the Hubble Space Telescope (HST). The Wide Field Camera 3 (WFC3) G141 grism data of this habitable zone ($T_{eq}$ = 235 K) Super-Earth (R = 1.7 $R_\oplus$), shows tentative evidence of water. However, the signal-to-noise ratio, and thus the significance of the detection, is low and stellar contamination models can cause modulation over the spectral band probed. We attempt to correct for contamination using these models and find that, while many still lead to evidence for water, some could provide reasonable fits to the data without the need for molecular absorption although most of these also cause features in the visible ground-based data which are nonphysical. Future observations with the James Webb Space Telescope (JWST) would be capable of confirming, or refuting, this atmospheric detection.

\vspace{7mm}
\end{abstract}

\section{Introduction}

Despite our strong observational bias for detecting large, gaseous giants -- similar to Saturn and Jupiter -- current statistics from over 4000 confirmed planets show a very different picture: planets roughly between 1--10\,$M_\oplus$ are the most abundant planets around other stars, especially around late-type stars \citep[e.g.][]{Dressing2013,Howard2016,Dressing2017,fulton2018}. 

Recent population statistics, backed up by theoretical models, reveal a surprising dichotomy in the occurrence rates of small planets. Precise radius measurements from the California-Kepler Survey (CKS), have indicated that they may come in two size regimes: Super-Earths with R$_{\rm p}$ $\le$ 1.5\,R$_{\oplus}$ and Sub-Neptunes with R$_{p}$\,=\, 2.0-3.0\,R$_{\oplus}$, with few planets in between \citep{Owen2017,fulton2017,fulton2018}. This natural division suggests that for planets larger than $1.8\,R_{\oplus}$, volatiles must contribute significantly to the planetary composition \citep{Rogers2010, Nettelmann2011, Valencia2013, Demory2011}, while smaller ones favour models with more negligible atmospheres \citep{Dressing2015, Gettel2016}. However, while various evolutionary models have postulated that this dichotomy is due to atmospheric loss, others have questioned it \citep[e.g.][]{zeng_ww} and only through atmospheric characterisation can this hypothesis be thoroughly tested.

The search for rocky planets with signs of habitability and bio-signatures form the holy grail of exoplanet atmospheric characterisation. Due to their larger relative size when compared to the host star, small planets around M-dwarfs have become the focus of this search. The TRAPPIST-1 system of 7 Earth-sized worlds \citep{gillon} provides some of the most intriguing targets for atmospheric characterisation. However, due to their lack of prominent features, HST observations of the four worlds which potentially lie within the habitable-zone of the star have ruled out the possibility of clear, hydrogen dominated atmosphere \citep{de_Wit_2016_trappist, de_Wit_2018_trappist}.

Thus far the smallest habitable-zone world with a confirmed water vapour detection is the 2.28\,R$_\oplus$, 7.96\,M$_\oplus$ planet K2-18\,b \citep{Tsiaras_k2-18, benneke_k2-18}. This detection has sparked intense debate regarding the nature of this world: water-world or Sub-Neptune? Seemingly sitting in the Sub-Neptune region of the radius valley, K2-18\,b's internal structure remains unknown. Large uncertainties on the radius of the star have caused the planet radius to be poorly defined, also affecting the calculated density. Hence, while the atmosphere of K2-18\,b could contain a large amount of hydrogen and/or helium, it could also be a water-world \citep{zeng_ww}. With current facilities, the search for atmospheric features of rocky, habitable-zone planets has not been successful thus far.

Here we present the analysis of Hubble WFC3 G141 observations of a temperate Super-Earth. With a radius of 1.7 R$_\oplus$ and a density of 7.5 gcm$^{-3}$, LHS\,1140\,b is likely to be a rocky world \citep{Ment_mass-radius_2019} and, with an equilibrium temperature of $\sim$235 K, is within the conservative habitable-zone of its star \citep{dittman_lhs1140b,kane_lhs_hz}. While recent ground-based observations were not precise enough to constrain atmospheric scenarios \citep{dia_lhs}, reconnaissance with Hubble WFC3 shows modulation in the transit depth over the 1.1-1.7 $\mu m$ wavelength range. We present atmospheric models that could fit this data but also show that stellar spot contamination can provide reasonable fits to the spectrum.

\section{Data Analysis}
\subsection{Reduction and Analysis of Hubble Data} 

Our analysis started from the raw spatially scanned spectroscopic images which were obtained from the Mikulski Archive for Space Telescopes\footnote{\url{https://archive.stsci.edu/hst/}}. Two transit observations of LHS\,1140\,b were acquired for proposal 14888 (PI: Jason Dittmann) and were taken in January and December 2017. Both visits utilised the GRISM256 aperture, and 256 x 256 subarray, with an exposure time of 103.13\,s which consisted of 16 up-the-ramp reads using the SPARS10 sequence. The visits had different scan rates with 0.10\,"/s and 0.14\,"/s used for January and December respectively, resulting in scan lengths of 10.9\," and 15.9\,".

We used Iraclis\footnote{\url{https://github.com/ucl-exoplanets/Iraclis}}, a specialised, open-source software for the analysis of WFC3 scanning observations \citep{tsiaras_hd209} and the reduction process included the following steps: zero-read subtraction, reference pixels correction, non-linearity correction, dark current subtraction, gain conversion, sky background subtraction, flat-field correction, and corrections for bad pixels and cosmic rays. For a detailed description of these steps, we refer the reader to the original Iraclis paper \citep{tsiaras_hd209}.

\begin{figure}
    \centering
    \includegraphics[width=\columnwidth]{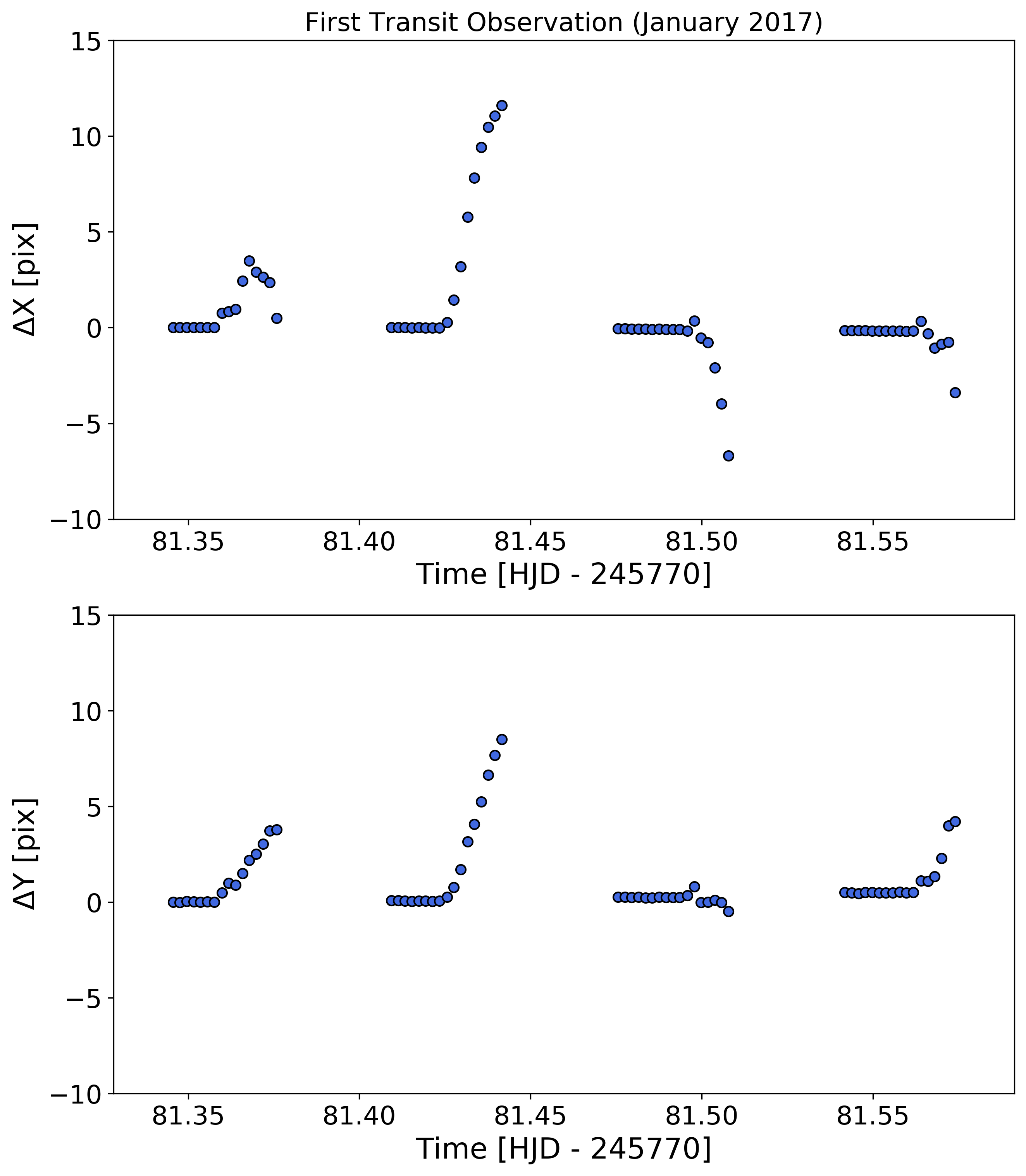}
    \includegraphics[width=\columnwidth]{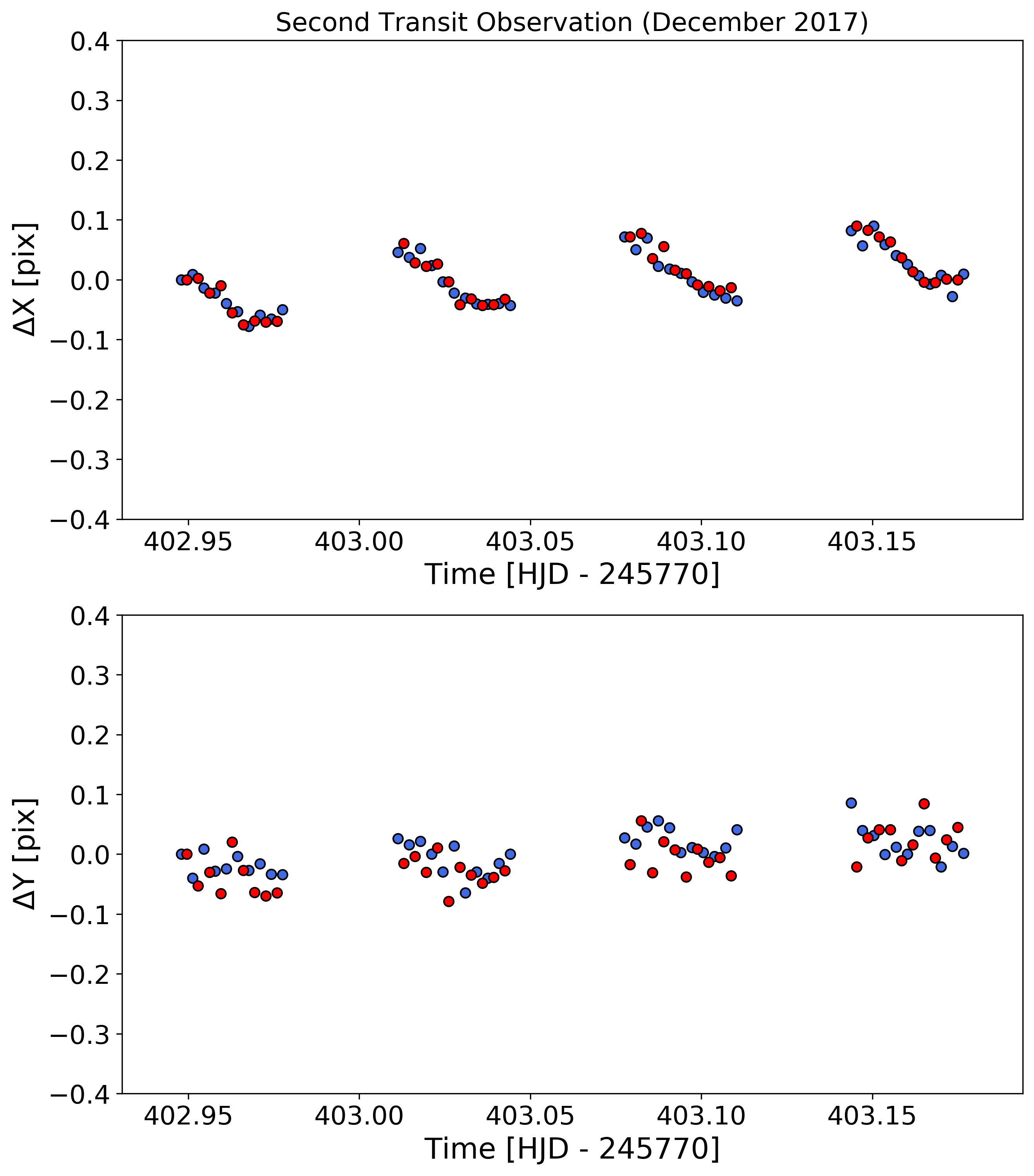}
    \caption{Shifts in the X and Y location of the spectrum for both observations. Black points indicate forward scans while reverse scans are shown in red.}
    \label{fig:shifts}
\end{figure}

\begin{figure}
    \centering
    \includegraphics[width = \columnwidth]{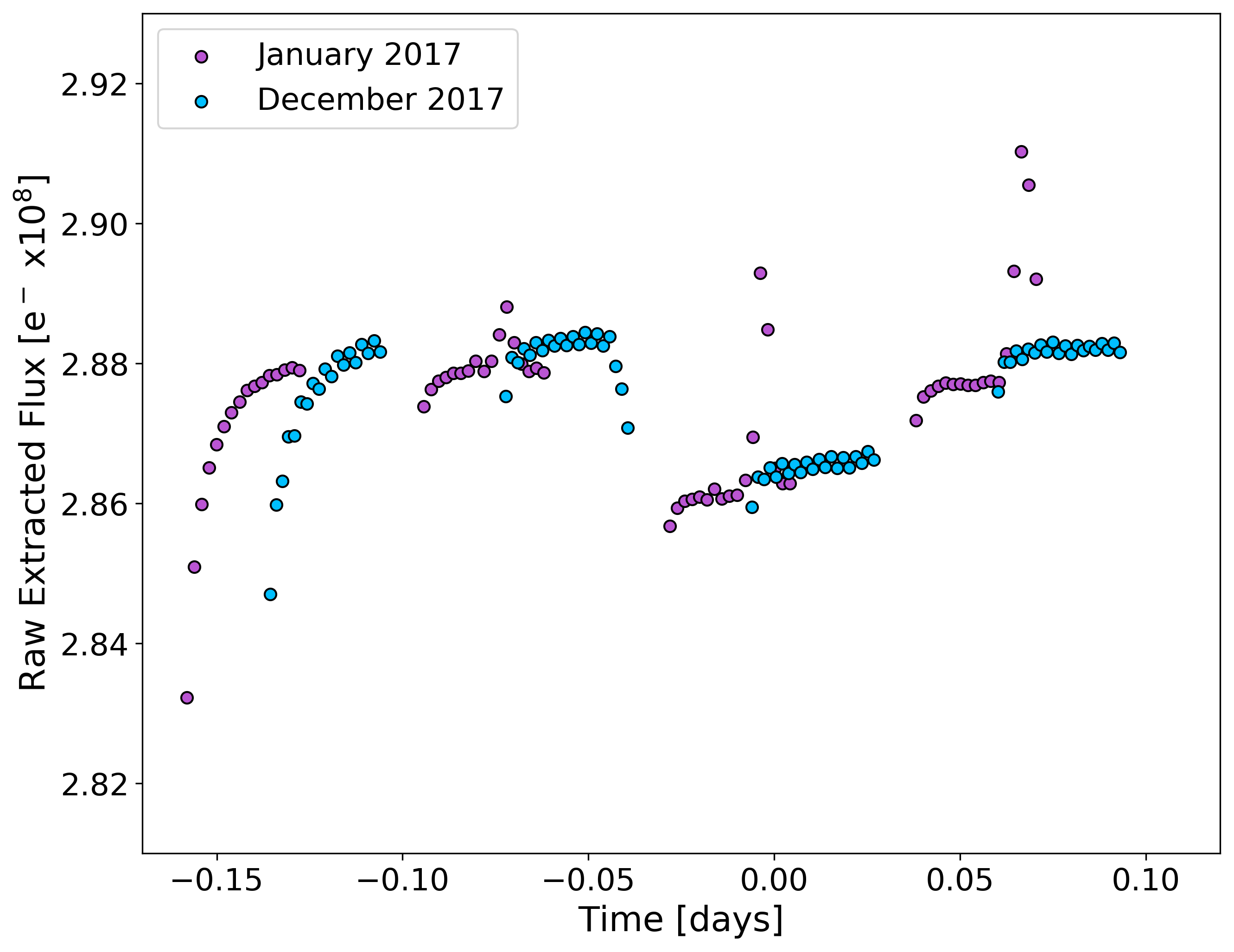}
    \caption{Raw extracted light curves for the LHS\,1140\,b observations. Even after removing obviously bad data from the January visit, a good fitting could not be achieved so the observation was discarded. We also note that the January visit did not include reverse scans while the December observation used both forward and reverse scans. Additionally, the visits had different scan rates and thus resulted in different scan lengths on the detector.}
    \label{fig:lhs_raw_lc}
\end{figure}

\begin{figure}
    \centering
    \includegraphics[width =\columnwidth]{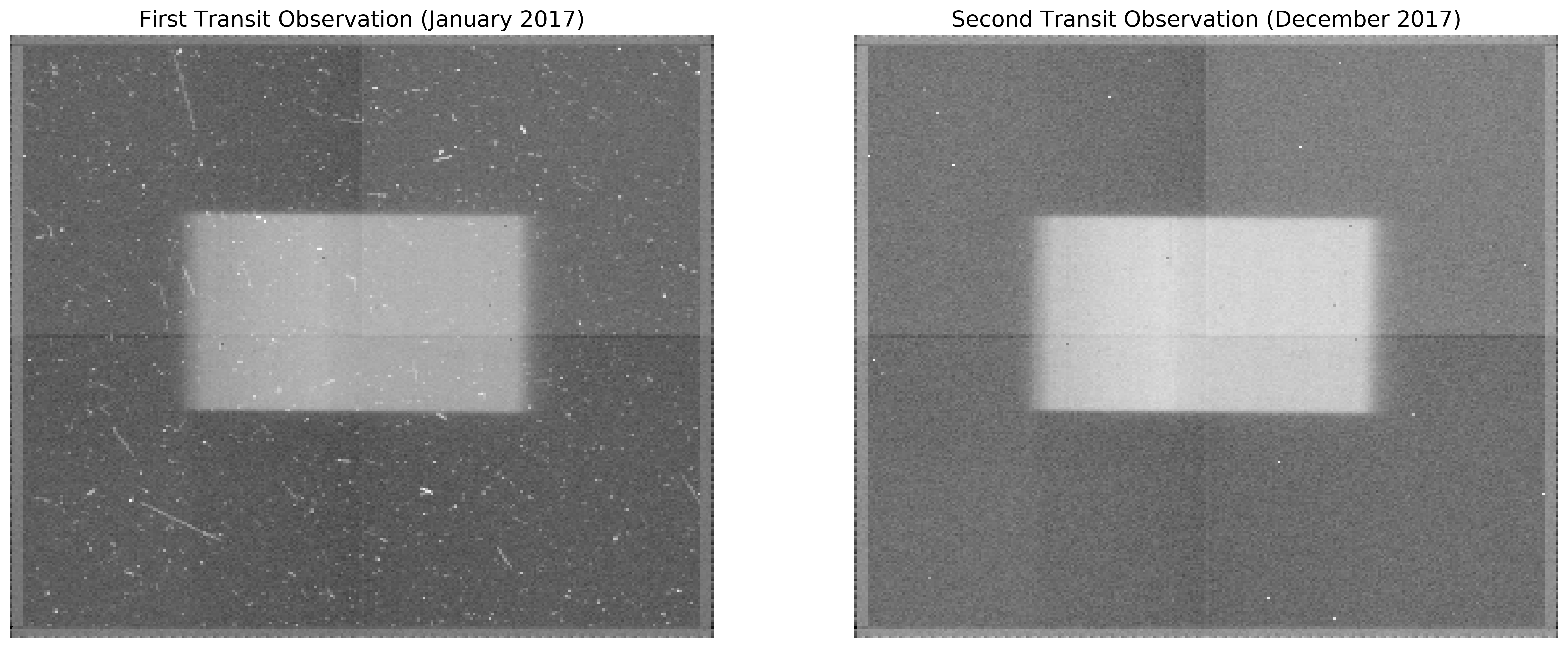}
    \caption{Example raw image from the first (left) and second (right) observations of LHS\,1140\,b from Proposal 14888. The degradation is clearly visible in the left hand image and was present for much of that visit.}
    \label{fig:bad_data}
\end{figure}

Although two transits of LHS 1140 b were obtained, one of these was affected due to large shifts in the location of the spectrum on the detector. These changes in position are shown in Figure \ref{fig:shifts} and, when the white light curve was extracted, large spikes were seen in the flux as shown in Figure \ref{fig:lhs_raw_lc}. The presence of such shifts are known to dominant the systematics and reduce precision if the position of the spectrum is not well-known \citep[e.g.][]{stevenson_shifts,tsiaras_systematics}. We attempted to remove the bad frames but still could not recover a satisfactory fit. Additionally, exposures that seemed to give reasonable data in the extracted white light curve also showed obvious degradation when the fits files were visually inspected, potentially due to an unusually high number of cosmic ray impacts. Figure \ref{fig:bad_data} displays an example raw image from each data set and the degradation is easily visible for the January observation. We therefore discarded the initial observation and so only the December data set was utilised.

For this observation, the reduced spatially scanned spectroscopic images were then used to extract the white (from 1.1-1.7 $\mu$m) and spectral light curves. The spectral light curves bands were selected such that the SNR is approximately uniform across the planetary spectrum. We then discarded the first orbit of each visit as they present stronger wavelength-dependent ramps, and the first exposure after each buffer dump as these contain significantly lower counts than subsequent exposures \citep[e.g.][]{deming_hd209, tsiaras_hd209}. 

We fitted the light curves using our transit model package PyLightcurve \citep{tsiaras_plc} which utilises the MCMC code ecmee \citep{emcee} and, for the fitting of the white light curve, the only free parameters were the mid-transit time and planet-to-star ratio. The other planet parameters were fixed to the values from \cite{Ment_mass-radius_2019} ($a/R_{*} = 95.34$, $i = 89.89^{\circ}$) while the limb darkening coefficients were computed using the formalism from \citet{claretI,claretII} and the stellar parameters from \citet{Ment_mass-radius_2019} ($T_{*} = 3216\,$K, $\log(g) = 5.0$).

\begin{table}
    \centering
    \begin{tabular}{cc} \hline \hline
    Parameter & Value \\ \hline \hline
    RA [J2000] & 00h 44min 59.3s \\
    Dec [J2000] & -15$^\circ$ 1' 18" \\
    $K_{\rm S}$ & 8.821 $\pm$ 0.024 \\
    $R_{\rm s}$ [$R_\odot$] & 0.2139 $\pm$ 0.0041\\
    $M_{\rm s}$ [$M_\odot$] & 0.179 $\pm$ 0.014 \\
    $T_{\rm s}$ [K] & 3216 $\pm$ 39 \\
    $\log(g)$ & 5.0 \\
    Fe/H & -0.24 $\pm$ 0.10 \\
    $R_{\rm p}$/$R_{\rm s}$ & 0.07390 $\pm$ 0.00008\\
    $M_{\rm p}$ [$M_{\oplus}$] & 6.98 $\pm$ 0.89\\
    $R_{\rm p}$ [$R_{\oplus}$] & 1.727 $\pm$ 0.032\\
    $\rho$ [ms$^{-2}$] & 7.5 $\pm$ 1.0 \\
    g [ms$^{-2}$] & 23.7 $\pm$ 2.7 \\
    $T_{\rm eff}$ [K] & 235 $\pm$ 5\\
    $S$ [$S_\oplus$] & 0.503 $\pm$ 0.030 \\
    $a$ [AU] & 0.0936 $\pm$ 0.0024\\
    $a/R_{\rm s}$ & 95.34 $\pm$ 1.06 \\
    $i$ [deg] & 89.89$^{+0.05}_{-0.03}$\\
    $e$ & $<$0.06$^*$ \\
    $P_{\rm orb}$ [days] & 24.7369148 $\pm$ 0.0000058$^\dagger$\\
    $T_{\rm mid}$ [BJD$_{\rm TDB}$] & 2457187.81760 $\pm$ 0.00012$^\dagger$\\ \hline
    $^*$Fixed to zero & $^\dagger$This work \\\hline \hline
    \end{tabular}
    \caption{Stellar and planetary parameters used or derived in this work. Data is from \citet{Ment_mass-radius_2019} unless otherwise stated.}
    \label{tab:para}
\end{table}

It is common for WFC3 exoplanet observations to be affected by two kinds of time-dependent systematics: the long-term and short-term `ramps'. These systematics were fitted using:

\begin{equation}
    R_w(t) = n^{scan}_w(1-r_a(t - T_0))(1-r_{b_{1}}e^{-r_{b_{2}}(t-t_{0})})
\end{equation}

\noindent where $t$ is time, $n^{scan}_w$ is a normalisation factor, $T_0$ is the mid-transit time, $t_o$ is the time when each HST orbit starts, $r_a$ is the slope of a linear systematic trend along each HST visit and ($r_{b_{1}},r_{b_{2}}$) are the coefficients of an exponential systematic trend along each HST orbit. The normalisation factor we used ($n^{scan}_w$) was changed to $n^{for}_w$ for upward scanning directions (forward scanning) and to $n^{rev}_w$) for downward scanning directions (reverse scanning). The reason for using different normalisation factors is the slightly different effective exposure time due to the known upstream/downstream effect \citep{McCullough_wfc3}.

{}

\begin{figure}
    \centering
    \includegraphics[width = \columnwidth]{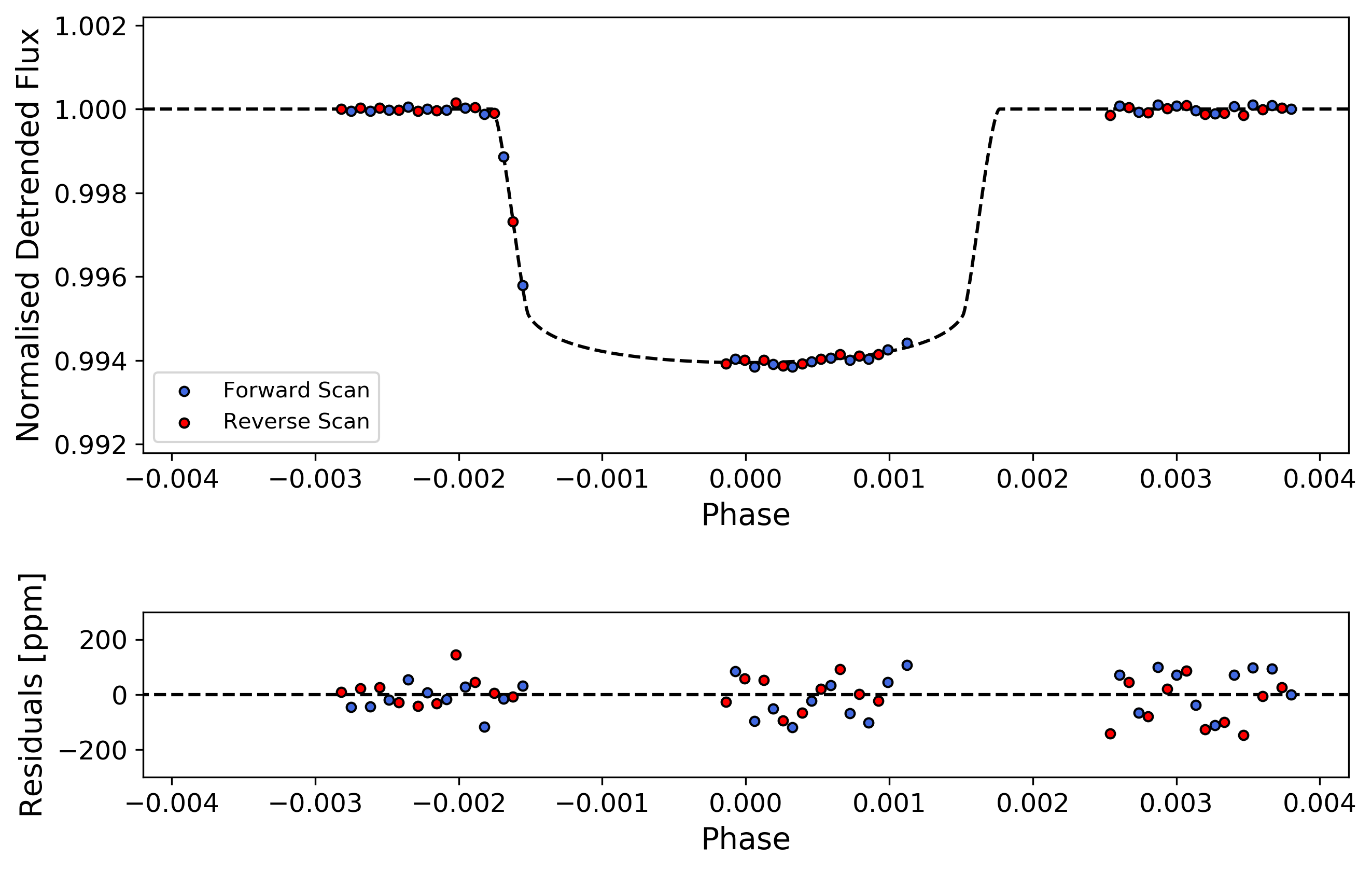}
    \caption{White light curve fit for December visit of LHS\,1140\,b. Top: detrended flux and best-fit model; Bottom: residuals from best-fit model.}
    \label{fig:lhs_hst_lc}
\end{figure}

We fitted the white light curve using the formulae above and the uncertainties per pixel, as propagated through the data reduction process. However, it is common in HST WFC3 data to have additional scatter that cannot be explained by the ramp model. For this reason, we scaled up the uncertainties in the individual data points, for their median to match the standard deviation of the residuals, and repeated the fitting, which is the standard practise for the Iraclis code \citep[e.g.][]{tsiaras_30planets,skaf_aresI}.

\begin{figure}
    \centering
    \includegraphics[width = \columnwidth]{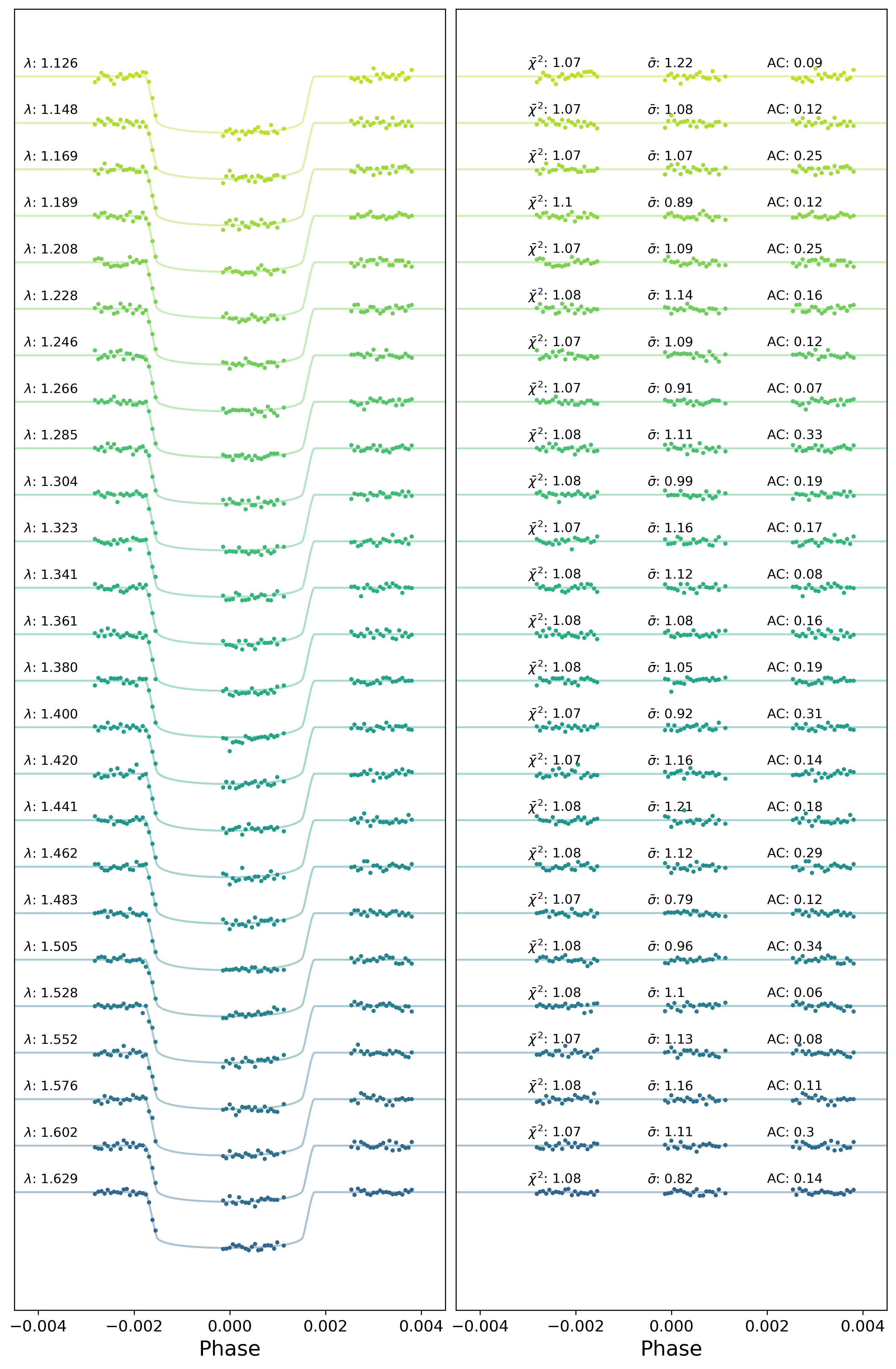}
    \caption{Spectral light curve fits from Iraclis for the transmission spectra of LHS\,1140\,b where, for clarity, an offset has been applied. In each plot, left panel: the detrended spectral light curves with best-fit model plotted; right panel: residuals from the fitting with values for the Chi-squared ($\chi^2$), the standard deviation with respect to the photon noise ($\bar{\sigma}$) and the auto-correlation (AC).}
    \label{fig:lhs_spec_lc}
\end{figure}{}

\begin{table*}[]
    \centering
    \begin{tabular}{ccccccccc} \hline \hline
    Wavelength [$\mu$m] & Transit Depth [\%] & Error [\%] & Bandwidth [$\mu$m] & $a_1$ & $a_2$ & $a_3$ & $a_4$ & Observatory \\ \hline
1.12625 & 0.5468 & 0.0087 & 0.0219 & 1.504 & -1.434 & 0.903 & -0.239 & HST \\
1.14775 & 0.5466 & 0.0077 & 0.0211 & 1.465 & -1.407 & 0.89 & -0.236 & HST\\
1.16860 & 0.5463 & 0.0075 & 0.0206 & 1.448 & -1.392 & 0.88 & -0.233 & HST\\
1.18880 & 0.5464 & 0.0064 & 0.0198 & 1.451 & -1.409 & 0.892 & -0.237 & HST\\
1.20835 & 0.5576 & 0.0080 & 0.0193 & 1.425 & -1.364 & 0.859 & -0.228 & HST\\
1.22750 & 0.5462 & 0.0077 & 0.0190 & 1.381 & -1.32 & 0.831 & -0.22 & HST\\
1.24645 & 0.5577 & 0.0078 & 0.0189 & 1.393 & -1.329 & 0.836 & -0.221 & HST\\
1.26550 & 0.5466 & 0.0060 & 0.0192 & 1.388 & -1.338 & 0.843 & -0.223 & HST\\
1.28475 & 0.5460 & 0.0070 & 0.0193 & 1.349 & -1.294 & 0.814 & -0.215 & HST\\
1.30380 & 0.5626 & 0.0070 & 0.0188 & 1.337 & -1.291 & 0.813 & -0.215 & HST\\
1.32260 & 0.5576 & 0.0082 & 0.0188 & 1.380 & -1.346 & 0.847 & -0.224 & HST\\
1.34145 & 0.5557 & 0.0079 & 0.0189 & 1.473 & -1.263 & 0.731 & -0.184 & HST\\
1.36050 & 0.5692 & 0.0079 & 0.0192 & 1.550 & -1.348 & 0.776 & -0.193 & HST\\
1.38005 & 0.5783 & 0.0079 & 0.0199 & 1.637 & -1.513 & 0.899 & -0.228 & HST\\
1.40000 & 0.5569 & 0.0068 & 0.0200 & 1.548 & -1.317 & 0.747 & -0.184 & HST\\
1.42015 & 0.5464 & 0.0080 & 0.0203 & 1.516 & -1.21 & 0.656 & -0.157 & HST\\
1.44060 & 0.5582 & 0.0089 & 0.0206 & 1.520 & -1.216 & 0.659 & -0.158 & HST\\
1.46150 & 0.546 & 0.0075 & 0.0212 & 1.495 & -1.195 & 0.651 & -0.157 & HST\\
1.48310 & 0.5548 & 0.0058 & 0.0220 & 1.499 & -1.199 & 0.650 & -0.156 & HST\\
1.50530 & 0.5349 & 0.0071 & 0.0224 & 1.525 & -1.266 & 0.697 & -0.169 & HST\\
1.52800 & 0.5463 & 0.0072 & 0.0230 & 1.505 & -1.254 & 0.696 & -0.170 & HST\\
1.55155 & 0.5576 & 0.0073 & 0.0241 & 1.484 & -1.244 & 0.694 & -0.170 & HST\\
1.57625 & 0.5470 & 0.0073 & 0.0253 & 1.502 & -1.296 & 0.727 & -0.178 & HST\\
1.60210 & 0.5460 & 0.0074 & 0.0264 & 1.493 & -1.322 & 0.751 & -0.185 & HST\\
1.62945 & 0.5462 & 0.0055 & 0.0283 & 1.481 & -1.371 & 0.801 & -0.200 & HST\\ \hline
1.38400 & 0.5520 & 0.0039 & 0.5920 & 1.463 & -1.302 & 0.766 & -0.194 & HST (White)\\\hline
0.8 & 0.5116 & 0.0334 & 0.4 & 3.222 & -4.915 & 4.330 & -1.451 & TESS \\\hline \hline
    \end{tabular}
    \caption{Recovered transit depths and associated errors for the HST and TESS data along with the limb-darkening coefficients used.}
    \label{tab:spectrum}
\end{table*}

Next, we fitted the spectral light curves with a transit model (with the planet-to-star radius ratio being the only free parameter) along with a model for the systematics ($R_\lambda$) that included the white light curve (divide-white method \citep{Kreidberg_GJ1214b_clouds}) and a wavelength-dependent, visit-long slope \citep{tsiaras_30planets} parameterised by:

\begin{equation}
    R_\lambda(t) = n^{scan}_\lambda(1-\chi_\lambda(t-T_0))\frac{LC_w}{M_w}
\end{equation}{}

\noindent where $\chi_\lambda$ is the slope of a wavelength-dependent linear systematic trend along each HST visit, $LC_w$ is the white light curve and $M_w$ is the best-fit model for the white light curve. Again, the normalisation factor we used, ($n^{scan}_\lambda$), was changed to ($n^{for}_\lambda$) for upward scanning directions (forward scanning) and to ($n^{rev}_\lambda$) for downward scanning directions (reverse scanning).

The white light curve fit is shown in Figure \ref{fig:lhs_hst_lc} and the subsequent spectral light-curve fits are shown in Figure \ref{fig:lhs_spec_lc}. A full list of stellar and planet parameters used in this study is given in Table \ref{tab:para} while the limb darkening coefficients and extracted spectrum are in Table \ref{tab:spectrum}.

\subsection{Atmospheric Modelling}

The retrieval of the transmission spectra was performed using the publicly available retrieval suite TauREx 3 \citep{al-refaie_taurex3}\footnote{\url{https://github.com/ucl-exoplanets/TauREx3_public}}. We included the molecular opacities from the ExoMol \citep{Tennyson_exomol}, HITRAN \citep{gordon} and HITEMP \citep{rothman} databases for: H$_2$O \citep{polyansky_h2o}, CH$_4$ \citep{exomol_ch4}, CO \citep{li_co_2015}, CO$_2$ \citep{rothman_hitremp_2010} and NH$_3$ \citep{ExoMol_NH3}. On top of this, we also included Collision Induced Absorption (CIA) from H$_2$-H$_2$ \citep{abel_h2-h2, fletcher_h2-h2} and H$_2$-He \citep{abel_h2-he} as well as Rayleigh scattering for all molecules. The priors used are listed in Table \ref{tab:priors}. We allowed the bounds on the volume mixing ratio (VMR) of each molecular species to vary from 1 to 1e$^{-12}$, allowing for both low and high mean molecular weight atmospheres. Our retrievals used 500 live points with an evidence tolerance of 0.5 and all retrievals for this study were performed on a single core of a 2017 MacBook Pro.

\subsection{Modelling of the effect of stellar spots}

\begin{figure}
    \centering
    \includegraphics[width = \columnwidth]{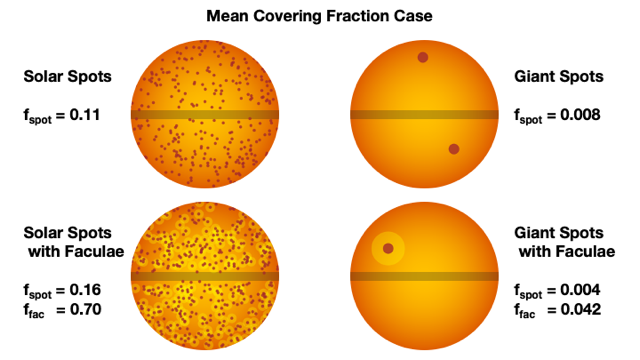}
    \includegraphics[width = \columnwidth]{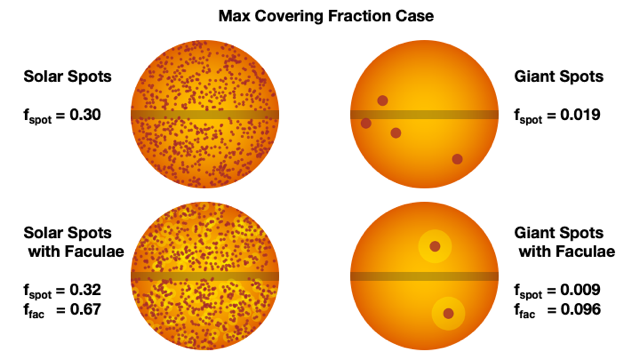}
    \caption{Graphical representation of the spot covering fractions considered in the work.}
    \label{fig:spots}
\end{figure}

We calculated the potential effect on transmission spectrum of the stellar spots using the model from \citet{Rackham_2018}. LHS\,1140 is known to have $\sim 1\%$ stellar brightness variability with a period of 131 days at optical wavelengths \citep{dittman_lhs1140b}. Assuming the variability is caused by stellar rotation, we infer the stellar surface is not homogeneous. We adopted four cases of the stellar spot distribution: giant spots, solar-type spots, giant spots with faculae, solar-type spots with faculae. These are depicted in Figure \ref{fig:spots} and the spot covering fraction values for each case are imported from the values for M4 stars in \citet{Rackham_2018}, as the stellar effective temperature and brightness variability is very similar to their values. 

In their paper, they calculate the total stellar flux by iteratively adding spots to random locations and derive the best spot covering fraction that represents a 1\% variation in brightness. Note that the amplitude of the brightness variability is not proportional to the spot covering fraction, since the size of each spot is defined ($R_{spot}$=$2^{\rm o}$ for solar-like spots and $R_{spot}=7^{\rm o}$ for giant spots), and multiple spots are distributed over the stellar surface in all cases. We assumed the photosphere to be at 3100 K, a spot temperature of 2700 K and faculae with temperatures of 3200 K. We used theoretical BT-Settl models of the stellar flux\footnote{\url{http://svo2.cab.inta-csic.es/theory/}} calculated for each temperature component, at the $\log g = 5$ and $\rm{[Fe/H]} = 0$. The effects on the transmission spectrum at each wavelength, the ``contamination factor", are calculated by the Equation 3 in \citet{Rackham_2018}. The derived contamination factor values are multiplied by a flat transit depth model, and we compared the atmospheric and stellar spots models to check which more adequately describes the observed transmission spectrum. 

\begin{table}
    \centering
    \begin{tabular}{cccc} \hline  \hline
    Parameters & \multicolumn{2}{c}{Prior Bounds} & Scale\\ \hline 
    $V_{x}$ & -12 & 0 & $\log_{10}$\\
    $T_{\rm term}$ [K] & 50 & 500 & linear \\
    $P_{\rm clouds}$ & 6 & -4 & $\log_{10}$ \\
    $R_{\rm p}$ [$R_{\rm jup}$] & 0.123 & 0.185 & linear \\ \hline \hline
    \end{tabular}
    \caption{List of the retrieved parameters, their uniform prior bounds, and the scaling used. The volume mixing ratios, denoted $V_{x}$ for a given molecule $x$, of all retrieved molecules were permitted to range up to 100\% of the atmospheric composition to search for evidence of a secondary atmosphere. Other parameters, such as the planet mass, were fixed to the values in Table \ref{tab:para}.}
    \label{tab:priors}
\end{table}

\subsection{TESS Data \& Ephemeris Refinement}

\begin{figure}
    \centering
    \includegraphics[width=\columnwidth]{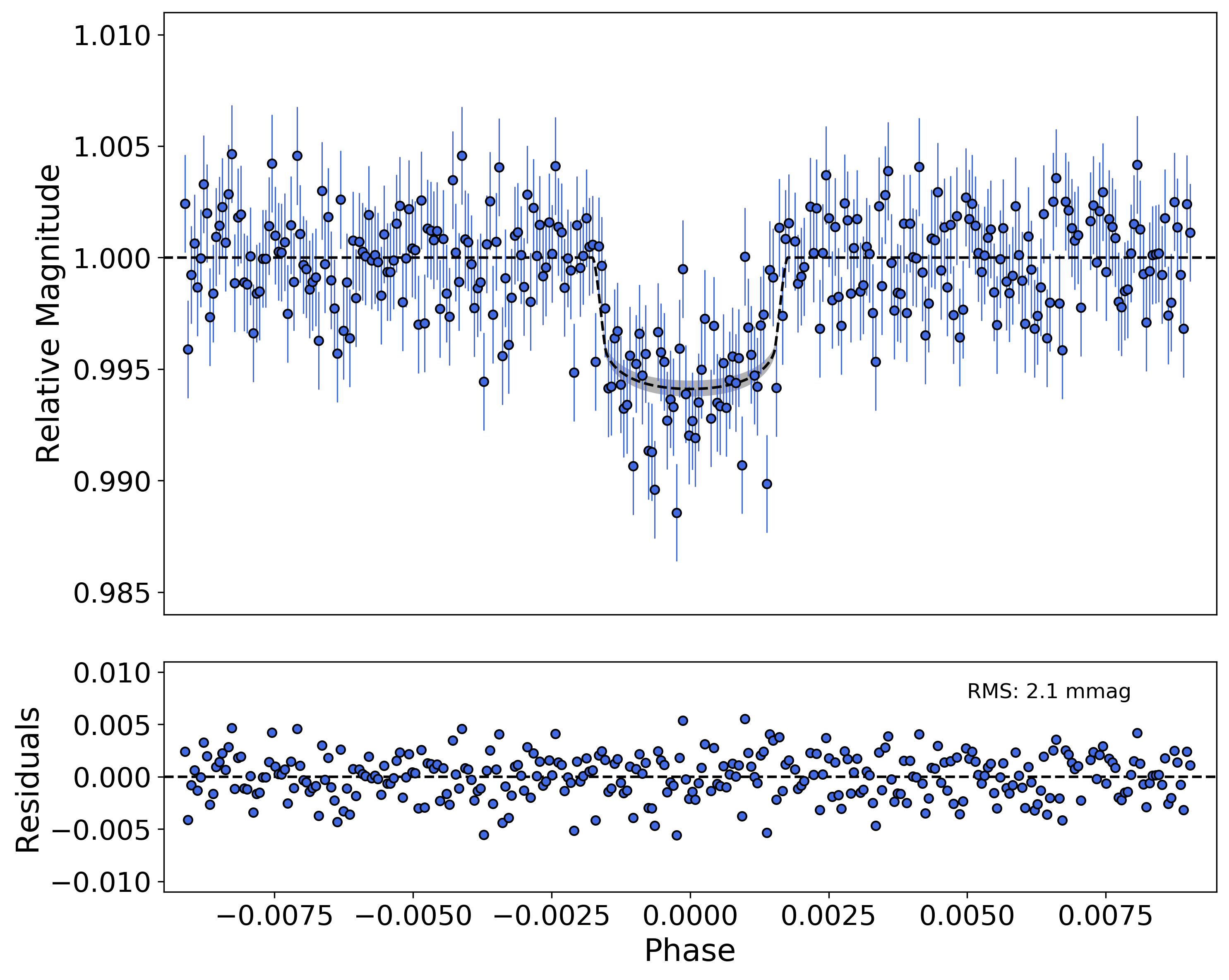}
    \caption{Fitting of TESS light curves from LHS\,1140\,b. Left: detrended light curves and best-fit model. Right: residuals from the fitting.}
    \label{fig:tess_lc}
\end{figure}

Accurate knowledge of exoplanet transit times is fundamental for atmospheric studies. To ensure that LHS\,1140\,b can be observed in the future, we used our HST white light curve mid time, along with data from TESS \citep{ricker}, to update the ephemeris of the planet. TESS data is publicly available through the MAST archive and we use the pipeline developed in \citet{edwards_orbyts} to download, clean and fit the 2 minute cadence Pre-search Data Conditioning (PDC) light curves \citep{smith_pdc,stumpe_pdc1,stumpe_pdc2}. LHS\,1140\,b had been studied in Sector 3 and, after excluding bad data, we recovered a single transit. Again we fitted only for the transit mid time and planet-to-star radius ratio, with all other values being fixed to those from \citet{Ment_mass-radius_2019}. For the limb darkening coefficients we utilise the values from \citet{claret_tess}. The extracted light curve is given in Table \ref{tab:tess_data} while the best-fit model is shown in Figure \ref{fig:tess_lc} and the mid time, which was used for refining the ephemeris, is given in Table \ref{tab:mid_times}. To get the period of the planet we fitted a linear function to the observations using a least-squared fit.

\begin{table}
    \centering
    \begin{tabular}{ccc} \hline \hline
    Epoch & Mid Time [BJD$_{\rm TDB}$] & Reference \\  \hline \hline
     -6 & 2456915.71154 $\pm$ 0.00004 & \citet{Ment_mass-radius_2019}\\
     42 & 2458103.083434 $\pm$ 0.000073 & This Work \\
     54 & 2458399.930786 $\pm$ 0.001305 & This Work \\ \hline \hline
    \end{tabular}
    \caption{Transit mid times of LHS\,1140\,b used in the ephemeris refinement.}
    \label{tab:mid_times}
\end{table}

\begin{table}[]
    \centering
    \begin{tabular}{ccc}\hline \hline
    Time [BJD$_{\rm TDB}$] & Normalised Flux & Error \\\hline
2458399.705274 & 1.002415 & 0.002193 \\
2458399.706663 & 0.995884 & 0.002188 \\
2458399.708051 & 0.999221 & 0.002191 \\
... & ... & ... \\
2458400.151098 & 0.996823 & 0.002189 \\
2458400.152487 & 1.002404 & 0.002194 \\
2458400.153876 & 1.001115 & 0.002191 \\ \hline \hline
    \end{tabular}
    \caption{Extracted TESS PDC light curve. The full table is available in a machine-readable format from the online journal. A portion is shown here for guidance.}
    \label{tab:tess_data}
\end{table}

\section{Results}

\subsection{Atmospheric Retrievals}

\begin{figure}
    \centering
    \includegraphics[width = \columnwidth]{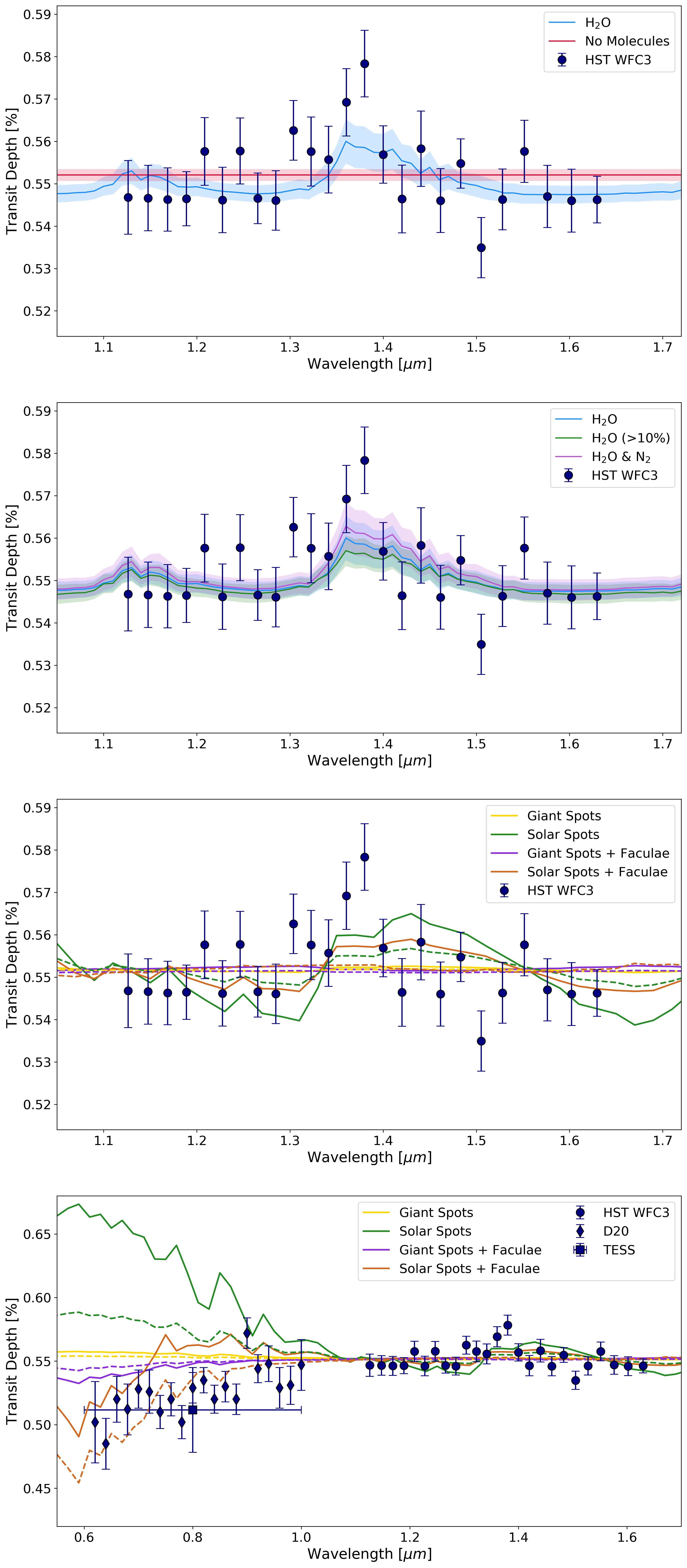}
    \caption{Best-fit models to the HST data from our atmosphere retrievals (top and upper middle) and stellar contamination models (lower middle). For the stellar contamination models, the solid lines depict the transit light source effect for maximum spot filling factor, as defined in \citet{Rackham_2018}, while the dashed lines represent the mean. Transit depths from \citet{dia_lhs} are also shown (bottom) and suggest the presence of solar spots with faculae.}
    \label{fig:best_fit}
\end{figure}

\begin{figure}
    \centering
    \includegraphics[width = \columnwidth]{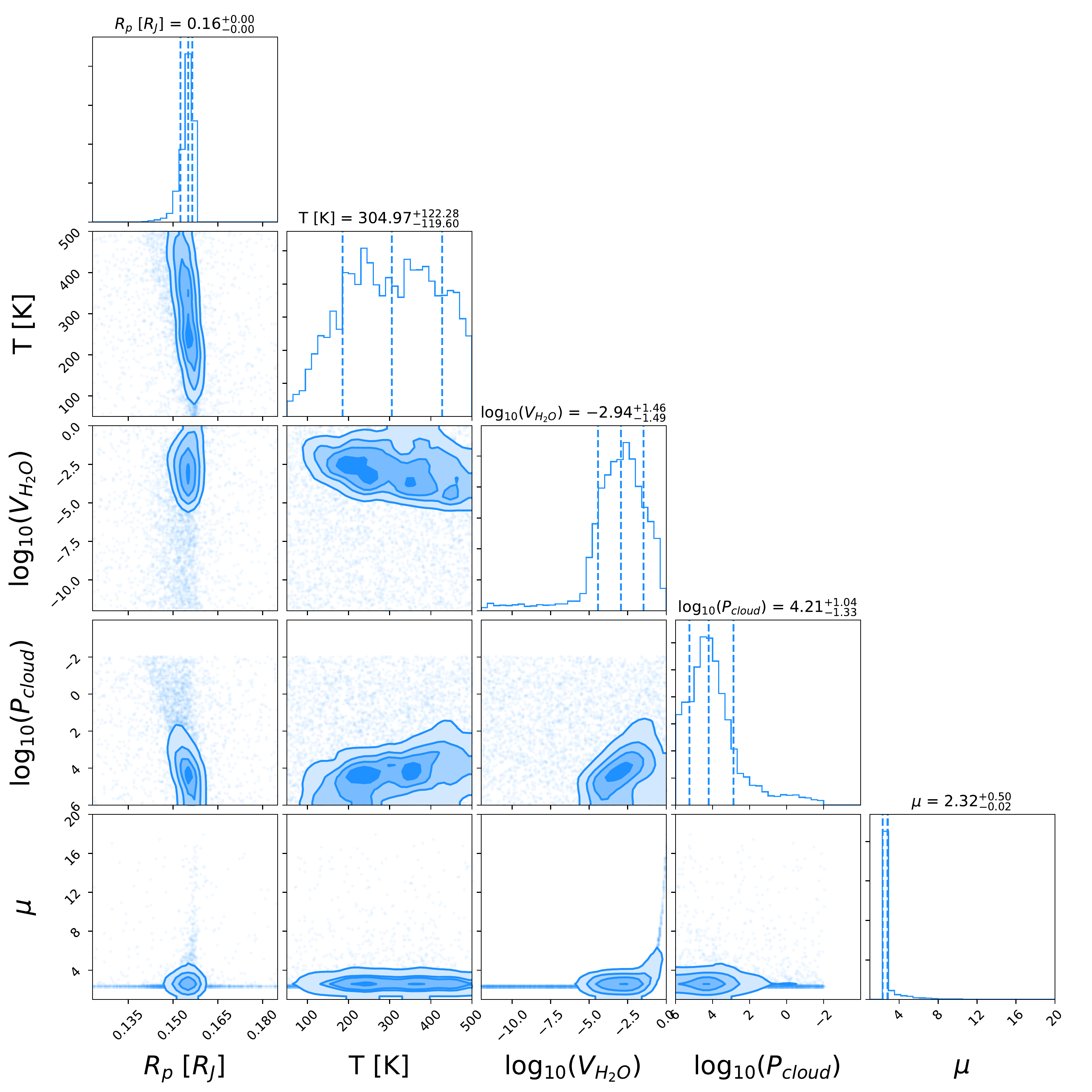}
    \caption{Posterior distributions for the water only atmospheric retrieval of LHS\,1140\,b. The water probability distribution is well-defined and the mean molecular weight indicates a primary atmosphere.}
    \label{fig:posteriors}
\end{figure}

The recovered spectrum is given in Table \ref{tab:spectrum} and while we ran atmospheric retrievals searching for a number of molecules, the only one for which the data supported any evidence for was H$_2$O. The best-fit spectrum is shown in Figure \ref{fig:best_fit} while Figure \ref{fig:posteriors} displays the posteriors of the H$_2$O only retrieval, which suggest an abundance of $\log_{10}(\text{V}_{\text{H}_2\text{O}})$ = -2.94$^{+1.45}_{-1.49}$. However, we note that the significance of the detection is relatively low. We compared the Bayesian log evidence (log(E)) for this retrieval to one which contained no molecular opacities. For this second retrieval the only fitted parameters were the planet radius, planet temperature and cloud-top pressure. Rayleigh scattering and CIA were also included. The difference in Bayesian log evidence was computed to be 2.26 in favour of the fit including H$_2$O, providing positive evidence for the detection of molecular features \citep{Kass1995bayes}. This is equivalent to the Atmospheric Detectability Index (ADI), as defined in \citet{tsiaras_30planets}, or 2.65$\sigma$. 

We note that the bounds used, which are given in Table \ref{tab:priors} allow for a higher mean molecular weight atmospheres, dominated by water, but the retrieval did not favour such a solution. Nevertheless, given the debate around the nature of such planets, we also attempted a retrieval which forced an atmosphere with a significant abundance of water ($> 10 \%$). In the baseline retrieval, the mean molecular weight was inferred to be 2.31$^{+0.28}_{-0.00}$ while this second case gives a value of 6.59$^{+5.73}_{-2.13}$ due to the volume mixing ratio of water being retrieved, to 1$\sigma$, as between 13.8 and 65.6\%. The forced retrieval does, in fact, give a marginally better fit to the data (log(E) = 195.21 versus log(E) = 194.47) and, compared to the flat model, is preferred by 2.95$\sigma$. However, given the small difference in evidence between the two retrievals including water, and that the Bayesian evidence is sensitive to the prior, one cannot use statistical means to preferentially select either \citep{Kass1995bayes}.


{}

{}

\begin{figure}
    \centering
    \includegraphics[width = \columnwidth]{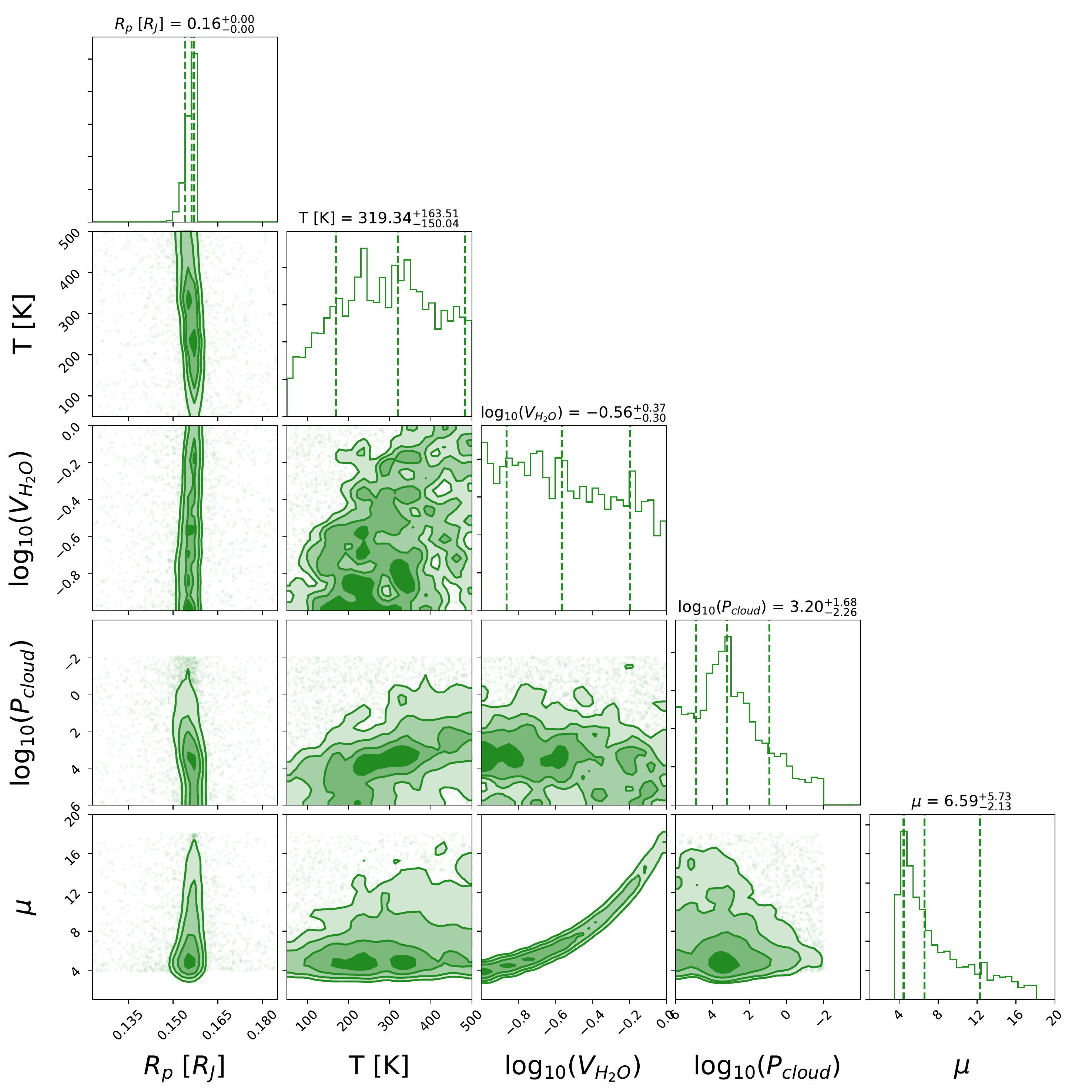}
    \caption{Posterior distributions for the water only atmospheric retrieval of LHS\,1140\,b where the water abundance is forced to be $>10\%$. The probability distribution of water is flat and the mean molecular weight highlights the shift towards a heavier atmosphere which is forced by this retrieval.}
    \label{fig:posteriors_heavy}
\end{figure}{}

\begin{figure}
    \centering
    \includegraphics[width = \columnwidth]{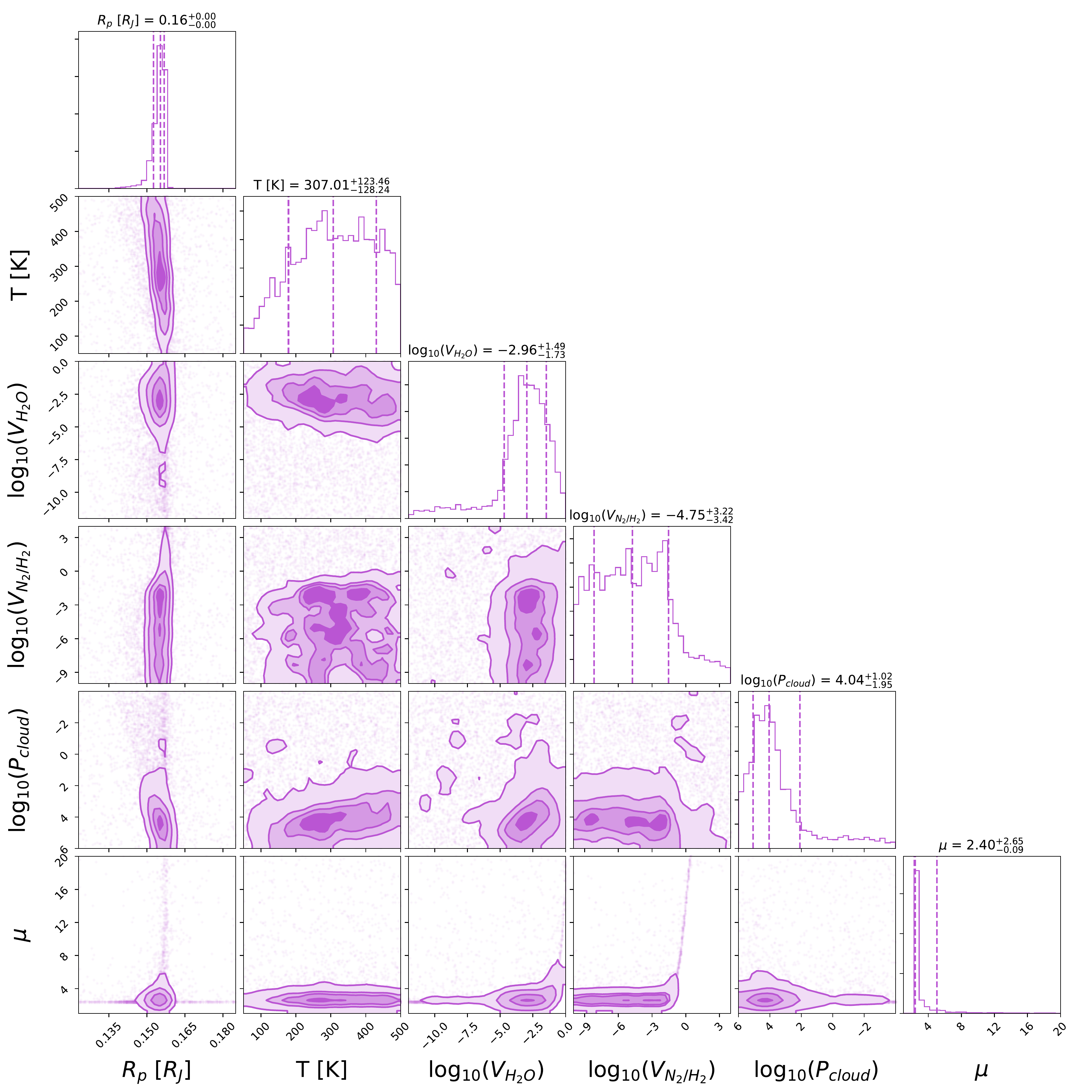}
    \caption{Posterior distributions fir the retrieval where, alongside the water abundance, the nitrogen to hydrogen ratio was fitted. The retrieved water abundance is highly similar to Figure \ref{fig:posteriors}.}
    \label{fig:posteriors_n2}
\end{figure}{}

As in \citet{Tsiaras_k2-18}, we also performed a retrieval which included nitrogen to increase the mean molecular weight without adding additional molecular absorption features. In this case, a water abundance of $\log_{10}(\text{V}_{\text{H}_2\text{O}})$ = -2.96$^{+1.49}_{-1.73}$ was recovered while the N$_2$/H$_2$ ratio was best-fit as -4.75 $^{+3.22}_{-3.42}$, as shown in Figure \ref{fig:posteriors_n2}. The Bayesian evidence (log(E) = 192.51) is again similar to the other models. Hence, while all our retrievals point to the presence of water the nature of the atmosphere (i.e. primary or secondary) cannot be ascertained. Additionally, while comparing the Bayesian evidence from different retrievals favours those with water, the different in evidence in all cases is small and it is worth noting that, compared to the flat model, the water only retrieval has but a single additionally fitting parameter ($\log_{10}(\text{V}_{\text{H}_2\text{O}})$). The water opacity adds great freedom to the model to fit the modulation without overly penalising the resulting Bayes factor for this increase dimensionality. For completeness, we also report the results of two retrievals with H$_2$O, NH$_3$, CH$_4$, CO and CO$_2$ as active absorbers: one where all molecules could form up to 100\% of the atmosphere and a second where all but water were capped at 10\% VMR. These resulted in a water abundances of $\log_{10}(\text{V}_{\text{H}_2\text{O}})$ = -3.31$^{+1.93}_{-4.11}$ and -3.74$^{+1.80}_{-4.39}$ respectively with evidences of log(E) = 194.51 and 194.33. Again positive evidence is found for atmospheric features but, in the latter case, the significance of the detection is reduced: the dimensionality of the fit has increased but the quality of it has not. Hence, while comparing the evidence from models is crucial, it must be done cautiously, with an understanding of the underlying statistical implications and the effects of the choice of one's priors: fine tuning one's priors can lead to apparent increases or decreases in the significance of a detection.

\subsection{Stellar Contamination}

The models of potential contamination are also plotted in Figure \ref{fig:best_fit}. For each, we compute the chi-squared as a means of comparing the ability of the model to fit the data. We additionally used the same metric to analyse the fit of the models from our retrievals. For simplicity, we chose to only compare the flat model and primary atmosphere containing water. As shown in Table \ref{tab:model_fit_hst}, the preferred case is still an atmosphere containing H$_2$O. We note that none of the star spot models alone provide convincing fits to the data, with the features induced coming at longer wavelengths and being broader than those seen in the data.

\begin{table}
    \centering
    \begin{tabular}{cccc}\hline\hline
    \multicolumn{2}{c}{Model} & $\chi^2$ & $ \bar{\chi}^2$ \\ \hline \hline
    \multirow{2}{*}{Atmosphere} & H$_2$O & 29.48 & 1.40\\
    & Flat & 35.82 & 1.63\\
    \multirow{2}{*}{Giant Spots} & mean & 35.43 & 1.61\\
    & max & 35.11 & 1.60\\
    \multirow{2}{*}{Solar Spots} & mean & 34.76 & 1.58\\
    & max & 51.83 & 2.36\\
    \multirow{2}{*}{Giant Spots + Faculae} & mean & 35.86 & 1.71\\
    & max & 36.05 & 1.72\\
    \multirow{2}{*}{Solar Spots + Faculae} & mean & 33.67 & 1.60\\
    & max & 34.18 & 1.63\\
    \hline \hline
    \end{tabular}
    \caption{Chi-squared ($\chi^2$) and reduced chi-squared ($\bar{\chi}^2$) values for different atmospheric and stellar contamination models for the HST WFC G141 data.}
    \label{tab:model_fit_hst}
\end{table}

Spectroscopic ground-based observations of LHS\,1140\,b were recently presented by \citet{dia_lhs}. They observed two transits of the planet and simultaneously monitored them with the IMACS and LDSS3C multi-object spectrographs on the twin Magellan telescopes. Their spectroscopic measurements resulted in a spectrum with a median error of 260 ppm which is significantly larger than the $\sim$70 ppm achieved here.

We note that the compatibility of different datasets is hard to confirm, particularly without spectral overlap, as the use of different limb darkening coefficients or orbital parameters, imperfect correction of instrument systematics, or stellar activity and stellar variability can all cause variations in the transit depth observed \citep[e.g.][]{tsiaras_30planets,alexoudi_inc,yip,yip_w96, changeat_k11, pluriel_aresIII}. \citet{dia_lhs} used the same orbital parameters, from \citet{Ment_mass-radius_2019}, but used logarithmic limb darkening coefficients. However, the TESS depth recovered here matches well with the ground-based data and was derived using the four-coefficient law from \citet{claret_tess}. In Figure \ref{fig:best_fit}, the data from \citet{dia_lhs} is plotted alongside the Hubble and TESS transit depths recovered here.

\begin{table}
    \centering
    \begin{tabular}{cccc}\hline\hline
    \multicolumn{4}{c}{No Re-normalisation} \\ \hline \hline
    \multicolumn{2}{c}{Model} & $\chi^2$ & $ \bar{\chi}^2$ \\ \hline \hline
    \multirow{2}{*}{Giant Spots} & mean & 95.43  & 5.61 \\
    & max & 107.45 & 6.32 \\
    \multirow{2}{*}{Solar Spots} & mean & 239.39 & 14.08 \\
    & max & 840.16 & 49.42 \\
    \multirow{2}{*}{Giant Spots + Faculae} & mean & 73.82 & 4.61  \\
    & max & 62.67 & 3.92 \\
    \multirow{2}{*}{Solar Spots + Faculae} & mean & 30.05 & 1.88 \\
    & max & 120.6 & 7.54 \\
    \hline \hline
    \multicolumn{4}{c}{With Re-normalisation} \\ \hline \hline
    \multirow{2}{*}{Giant Spots} & mean & 33.8 & 1.99 \\
    & max & 36.0 & 2.12  \\
    \multirow{2}{*}{Solar Spots} & mean & 61.11 & 3.59 \\
    & max & 186.8 & 10.99\\
    \multirow{2}{*}{Giant Spots + Faculae} & mean & 29.15 & 1.82 \\
    & max & 25.74 & 1.61 \\
    \multirow{2}{*}{Solar Spots + Faculae} & mean & 28.09 & 1.76 \\
    & max & 34.1 & 2.13 \\
    \hline \hline
    \end{tabular}
    \caption{Chi-squared values for different atmospheric and stellar contamination models for the data from \citet{dia_lhs}.}
    \label{tab:model_fit_ground}
\end{table}

\begin{figure}
    \centering
    \includegraphics[width = \columnwidth]{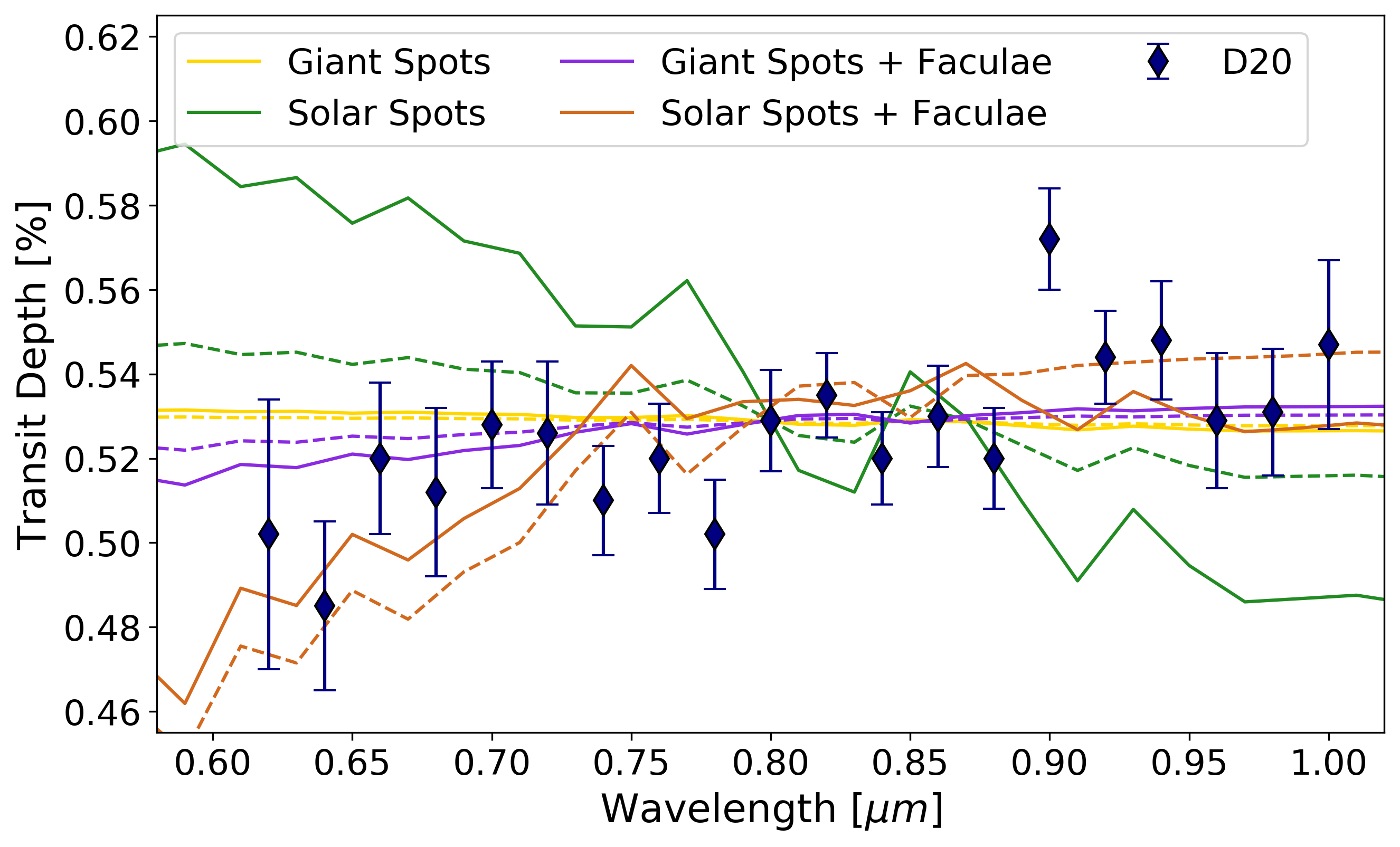}
    \caption{Stellar contamination models when re-normalised to the ground-based data from \citet{dia_lhs}.}
    \label{fig:ground_cont}
\end{figure}

\begin{table*}[]
    \centering
    \begin{tabular}{ccccccc}\hline \hline
     \multicolumn{2}{c}{Stellar Model} & Water Abundance & log(E) Water & log(E) Flat & $\Delta$log(E) & Sigma \\ \hline \hline
     \multirow{2}{*}{Giant Spots} & mean & -2.85$^{+1.48}_{-1.69}$ & 194.07 & 192.43 & 1.64 & 2.37 \\
                                   & max & -2.98$^{+1.68}_{-1.69}$ & 194.17 & 192.51 & 1.66 & 2.38 \\
     \multirow{2}{*}{Solar Spots} & mean & -4.31$^{+2.57}_{-4.75}$ & 193.12 & 192.71 & 0.41 & 0.26 \\
                                   & max & -7.11$^{+3.41}_{-3.32}$ & 184.30 & 184.50 & -0.20 & - \\
     \multirow{2}{*}{Giant Spots + Faculae} & mean & -2.92$^{+1.51}_{-1.42}$ & 194.50 & 192.07 & 2.43 & 2.73 \\
                                   & max & -2.78$^{+1.36}_{-1.41}$ & 194.58 & 192.04 & 2.54 & 2.77 \\
                                   
     \multirow{2}{*}{Solar Spots + Faculae} & mean & -2.78$^{+1.43}_{-1.70}$ & 195.37 & 193.24 & 2.13 & 2.60  \\
                                   & max & -5.57$^{+3.20}_{-4.27}$ & 193.57 & 193.11 & 0.46 & 0.26 \\ \hline            
    \multicolumn{2}{c}{None} & -2.94$^{+1.45}_{-1.49}$ & 194.47 & 192.21 & 2.26 & 2.65 \\ \hline \hline
    \end{tabular}
    \caption{Retrieved water abundance after accounting for different stellar contamination models. The Bayesian evidence for the models with and without water are also given along with the sigma level for each.}
    \label{tab:remove_cont}
\end{table*}

The ground-based dataset appears to show a shallower transit depth than any of the stellar contamination models but the solar spots with faculae model provides the best-fit as demonstrated by the chi-squared values in Table \ref{tab:model_fit_ground}. The plotted data in Figure \ref{fig:best_fit} assumes no offsets between datasets but, to ensure we don't draw false conclusions because of one, we re-normalise the stellar contamination models. To do this we sample various offsets for each stellar model, finding the best-fit value to the ground-based data alone using the chi-squared metric. These are shown in Figure \ref{fig:ground_cont} and the chi-square values are again given in Table \ref{tab:model_fit_ground}.

Given the downward slope seen in this dataset, this would appear to rule out the case of solar spots without facuale, which would provide the largest modulation in the HST wavelength range. Of the stellar models computed, before re-normalisation the mean coverage of solar spots with facuale provides the best-fit to the slope seen in this ground-based dataset: the same model also best fits the HST data. After re-normalisation, solar spots and faculae in both coverage cases give a reasonable fit. Additionally, giant spots and facuale, with either mean or max coverage, also provide good fits to the IMACS/LDSS3C dataset and these do not cause modulation in the WFC3 bandpass.

\subsection{Impact of Accounting for Stellar Contamination}

Of course, one would not expect the spectrum to be explained entirely by the stellar model unless LHS\,1140\,b were to have no atmosphere. Hence, the signal seen should be a combination of contributions. To explore the effect of stellar contamination on the retrieved atmosphere, and thus the significance of the water detection, we tested subtracting our spot models from both our HST data set and the observations from \citet{dia_lhs}. The effect of stellar contamination is greater in the visible. Hence, we use this to rule out spot contamination models which would imply nonphysical atmospheric features in the data from \citet{dia_lhs}. A similar approach was taken by \citet{wakeford_trap} for TRAPPIST-1\,g. While a slope which increases at shorter wavelengths can be described by Rayleigh scattering, a negative one would be more difficult to describe. If solar spots were the correct contamination model, the atmospheric feature would be around 40 scale heights in the max coverage case, or 30 in the mean, and thus is not feasible. Other models still have large feature sizes, 10-20 scale heights, but the data is relatively noisy and the error bars are of the order of several scale heights.


\begin{figure*}
    \centering
    \includegraphics[width = 0.45\textwidth]{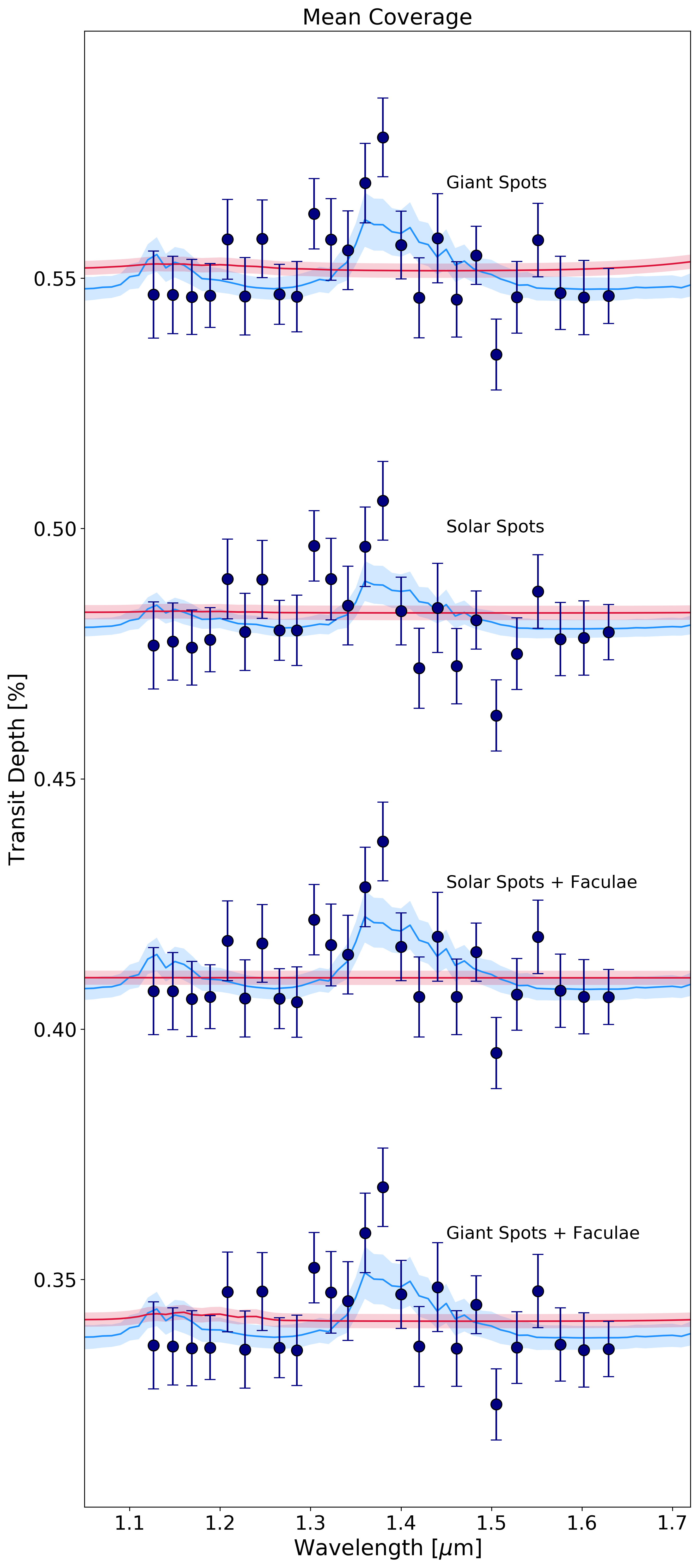}
    \includegraphics[width = 0.45\textwidth]{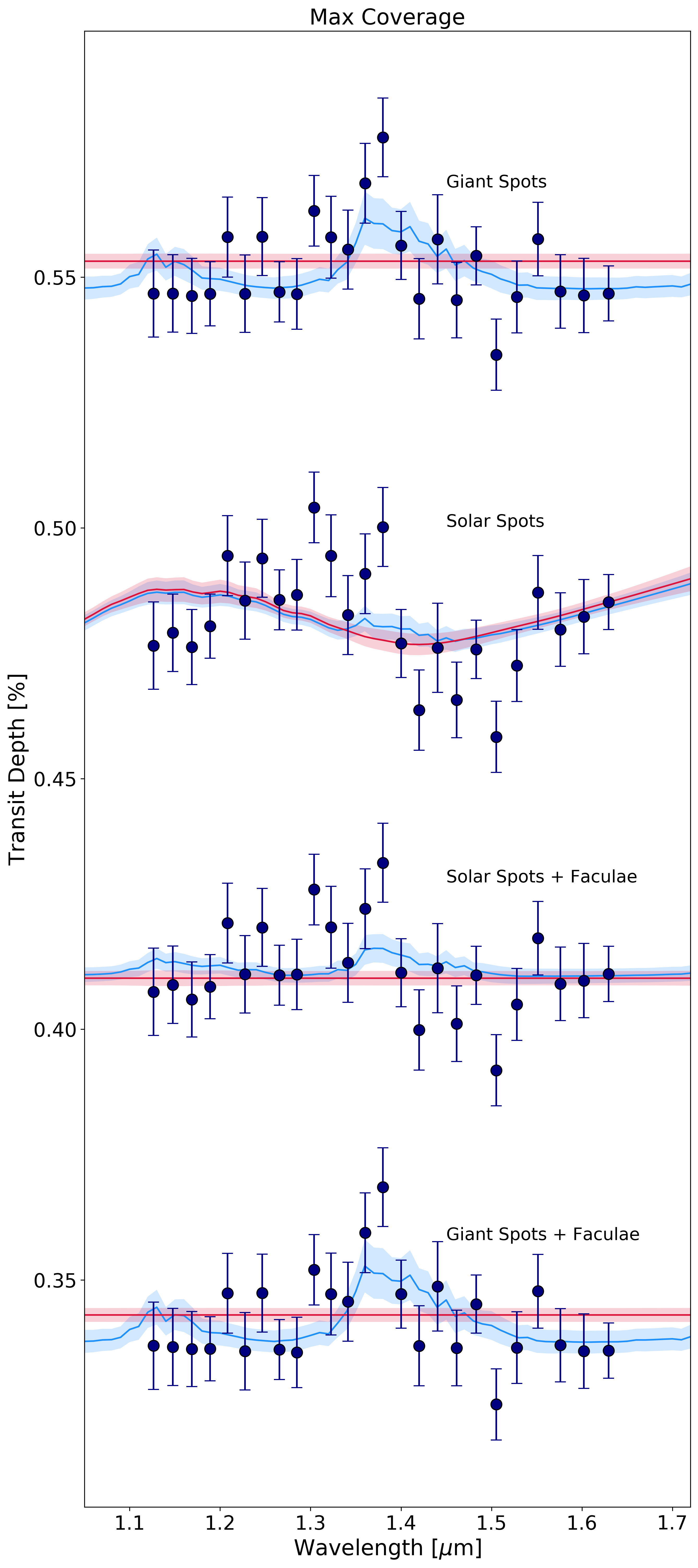}
    \caption{Corrected HST WFC3 spectrum for different stellar contamination models and the subsequent best-fits from retrievals with (blue) and without (red) water. In the majority of cases, a solution with water is preferred. The modulation seen in the retrievals on the data corrected for max coverage of solar spots is due to CIA.}
    \label{fig:hst_after_corr}
\end{figure*}

For the HST data, we also accounted for each stellar contamination model by subtracting the contribution and conducting two retrievals on the resultant spectrum: one with water and one without. The only three models which cause significant modulation in the HST WFC3 band pass are solar spots, both mean and max coverage, and the max coverage of solar spots with faculae. Subtracting each of these essentially completely removed any evidence for water in the atmosphere. However, as mentioned, the visible data could not be explained in solar spots case. Plots of the subsequent HST WFC3 data, with best-fit models, are given in Figure \ref{fig:hst_after_corr}. The ``corrected'' spectrum in the max solar spots case is strange and the retrieval tries to use CIA to explain the modulation. The maximum coverage of solar spots and faculae is the only stellar contamination model which leads to no evidence for water in the atmosphere of LHS\,1140\,b and is not ruled out by the visible data.


Having accounted for all other models, our retrievals still favoured the presence of water with a confidence of 2.38-2.77$\sigma$ and with an abundance similar to that from our baseline retrieval on the unmodified HST data. 

\subsection{Impact of Removing Data Points or Utilising Different Fitting Parameters}

The spectrum obtained using HST WFC3 contains 25 data points but the evidence for water is likely to be driven by only a few of these: those around 1.4 $\mu$m where the feature is the strongest. Therefore we attempted retrievals on data sets where we removed individual data points. Each time we ran the model with and without water and compared the difference in the global log evidence. Our analysis found that removing the data point at 1.38 $\mu$m eliminates all indications from water being present with the removal of the 1.36 $\mu$m data point reducing the confidence to 2.01$\sigma$. Meanwhile, deducting the spectral data at 1.40 or 1.42 $\mu$m changes the confidence of the models that water is present to 2.57 and 2.9 $\sigma$ respectively. Such results are expected given the narrow wavelength region means only one water feature is probed.

\begin{figure}
    \centering
    \includegraphics[width = \columnwidth]{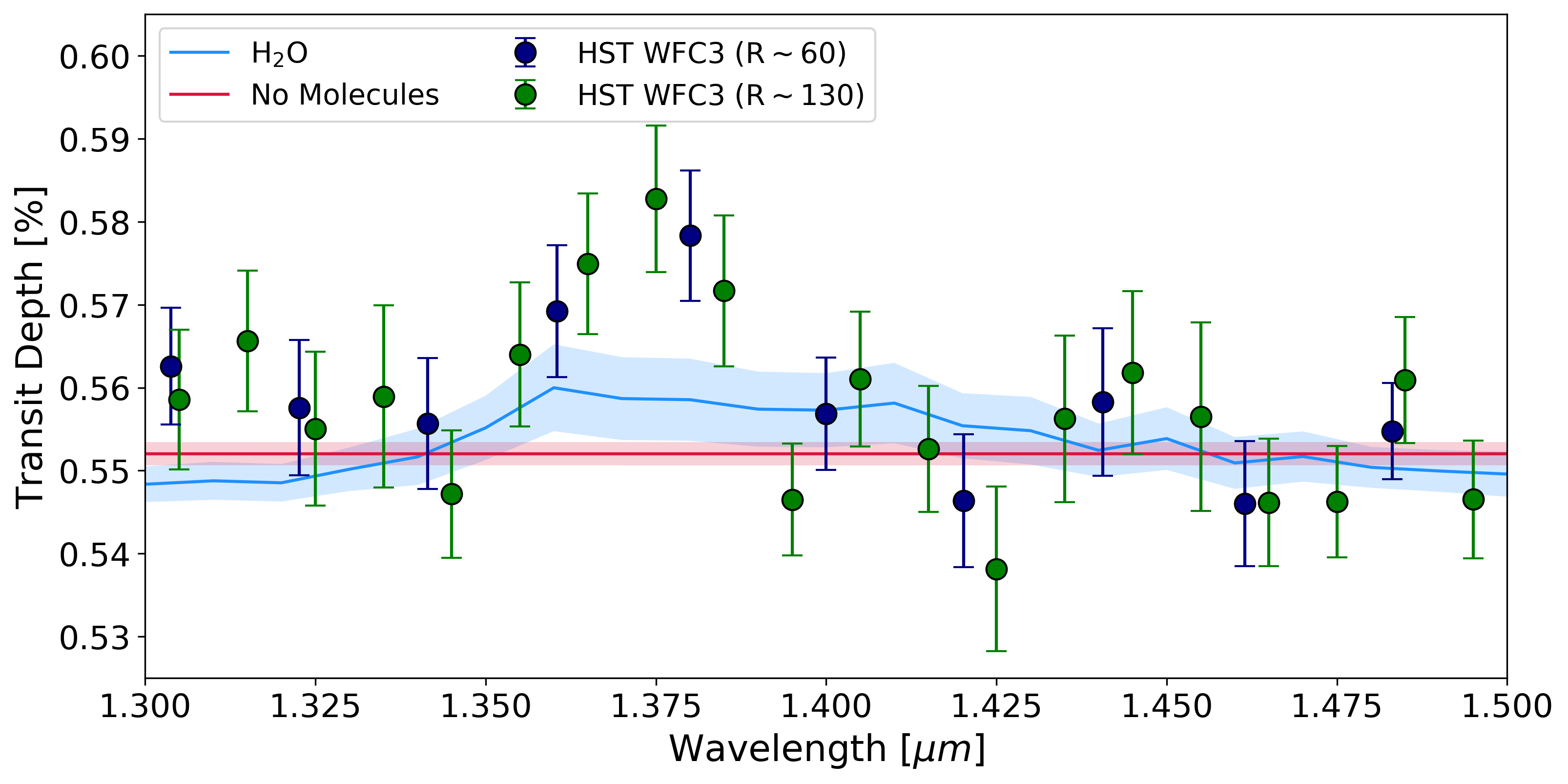}
    \caption{Higher resolution HST WFC3 G141 spectrum plotted alongside the main spectral dataset and best-fit models. The high resolution spectrum also peaks around the water feature at 1.4 $\mu$m and several data points form this feature, suggesting it is not caused by narrow-band contamination.}
    \label{fig:high_res}
\end{figure}
\begin{figure}
    \centering
    \includegraphics[width = \columnwidth]{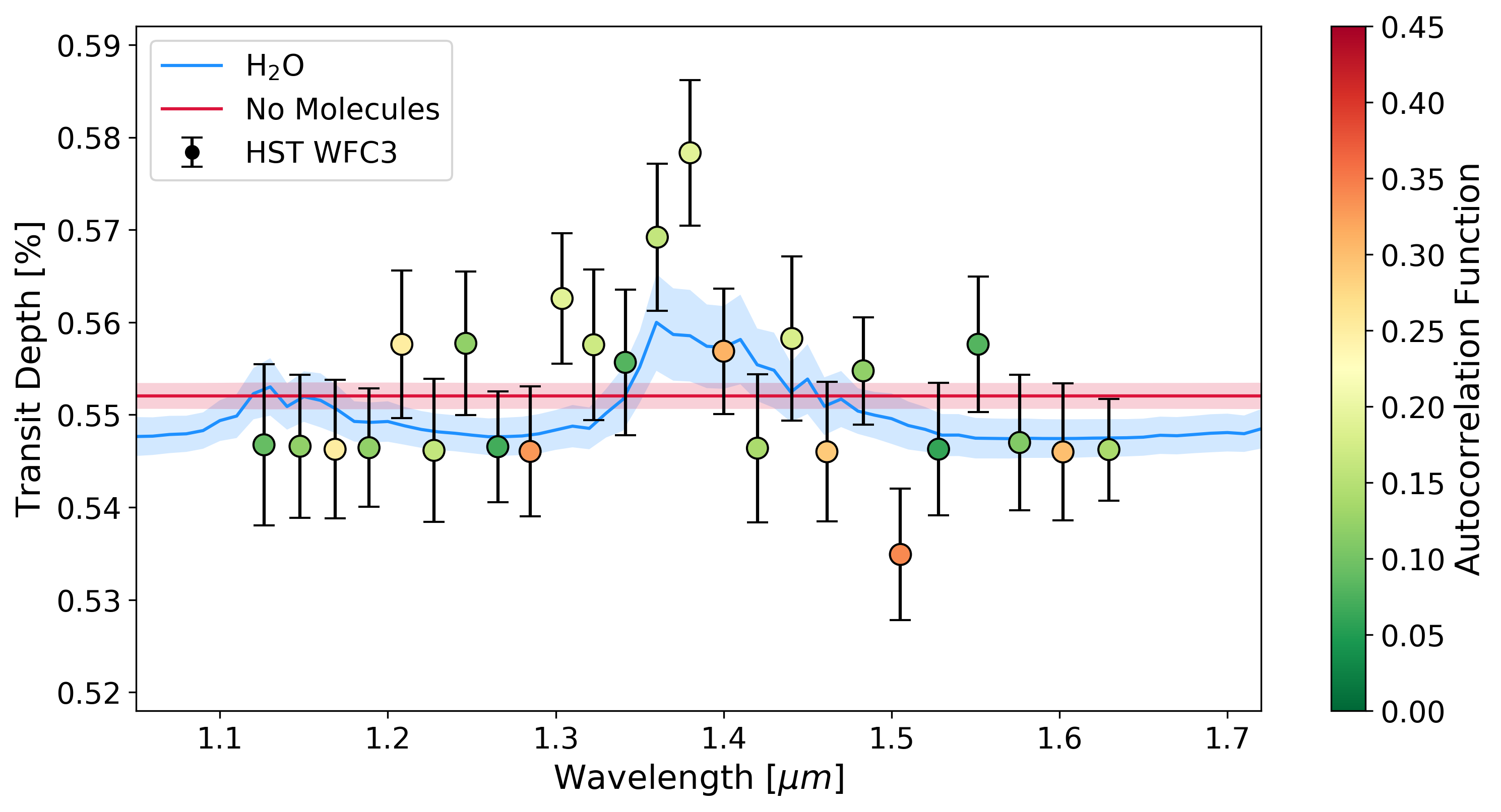}
    \caption{HST WFC3 G141 with the data points coloured by value of the auto-correlation function for their fitting.}
    \label{fig:autocor}
\end{figure}

To further explore the quality of the spectral light curve fitting, and thus of the water detection, we also produced a spectrum at the native resolution of the G141 grism (R$\sim$130 at 1.4 $\mu$m). A peak of several data points is seen in this dataset around the 1.4 $\mu$m water feature, as seen in Figure \ref{fig:high_res}, suggesting the peak seen is not due to a narrow band contamination of the spectra. Additionally, we studied the auto-correlation function of each spectral light curve fit. Various correlations are calculated (e.g. between one point and the next) using the numpy.correlate package and the maximum value is reported. Figure \ref{fig:autocor} shows that the 1.38 $\mu$m data point, the one on which the water detection hinges, appears to be well fitted and thus reliable.

\begin{figure}
    \centering
    \includegraphics[width = \columnwidth]{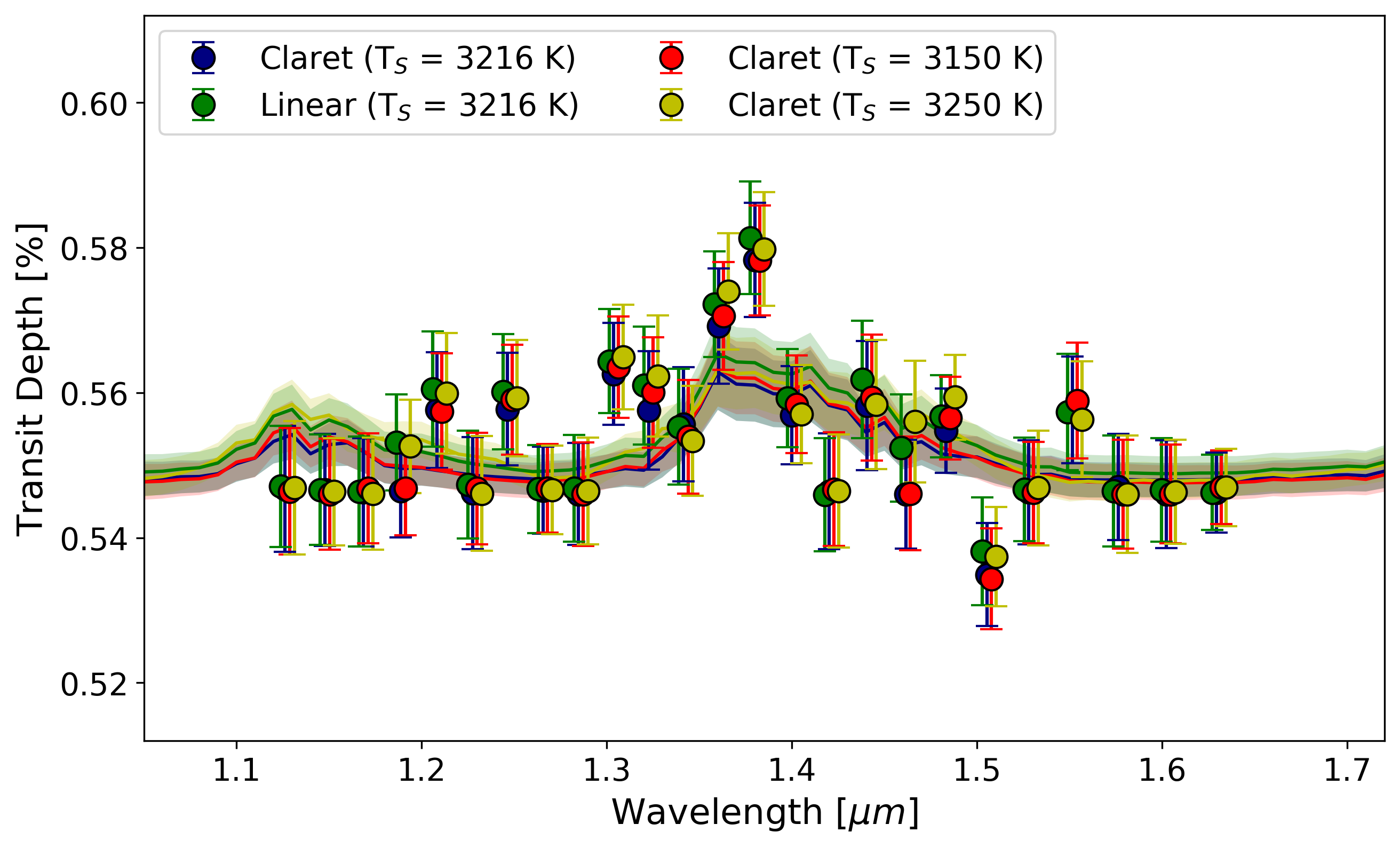}
    \caption{The spectra recovered from the fitting using different limb darkening coefficients. They are all within 1$\sigma$ of each other and, from our retrievals, prefer the presence of water to a flat model.}
    \label{fig:extra_ldc}
\end{figure}

\begin{figure}
    \centering
    \includegraphics[width = \columnwidth]{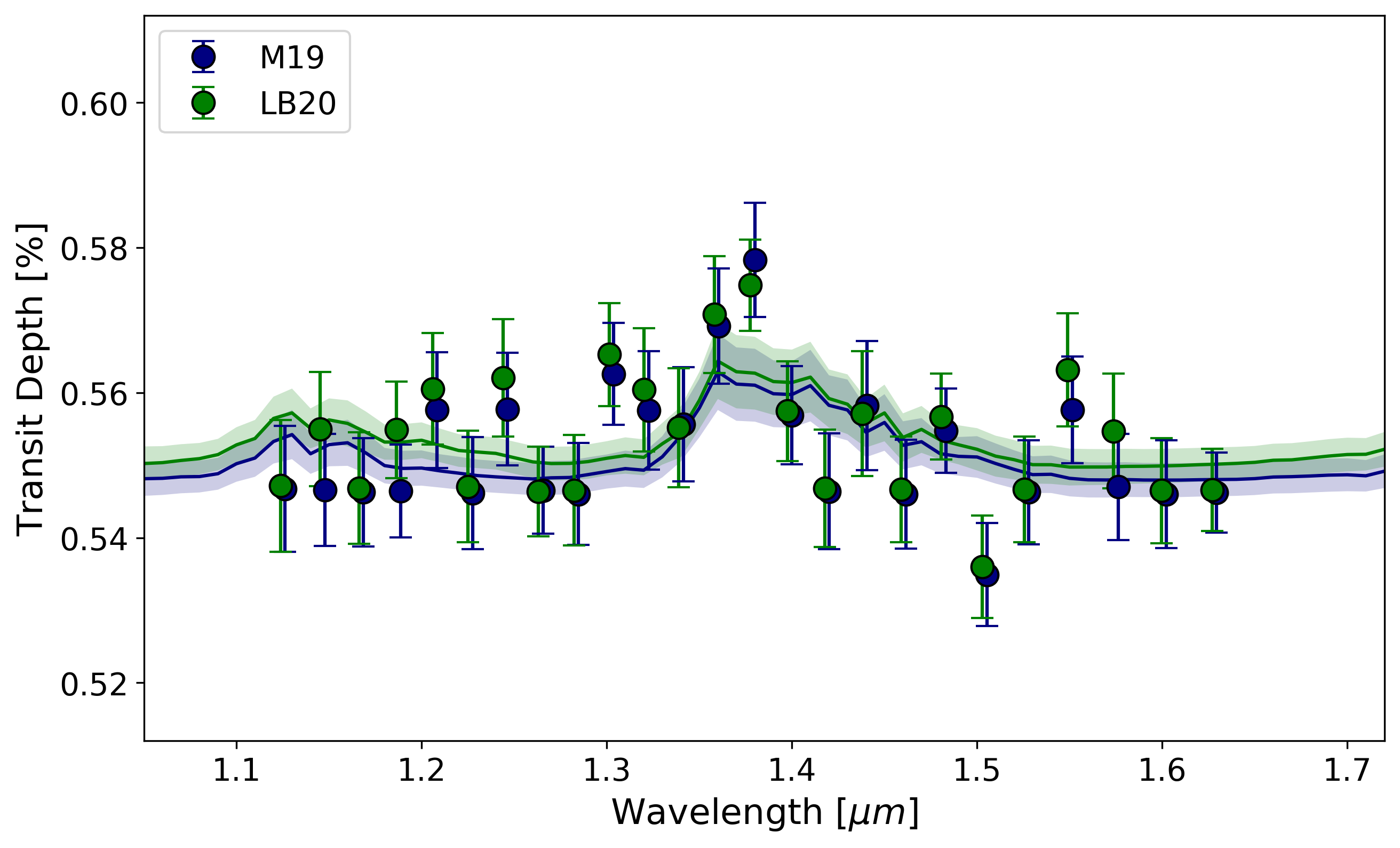}
    \caption{The spectra recovered from the fitting using different parameters from \citep{Ment_mass-radius_2019} and \citep{lillo_lhs_water}. They are within 1$\sigma$ of each other and, from our retrievals, prefer the presence of water to a flat model.}
    \label{fig:new_para}
\end{figure}

\begin{figure*}
    \centering
    \includegraphics[width = \textwidth]{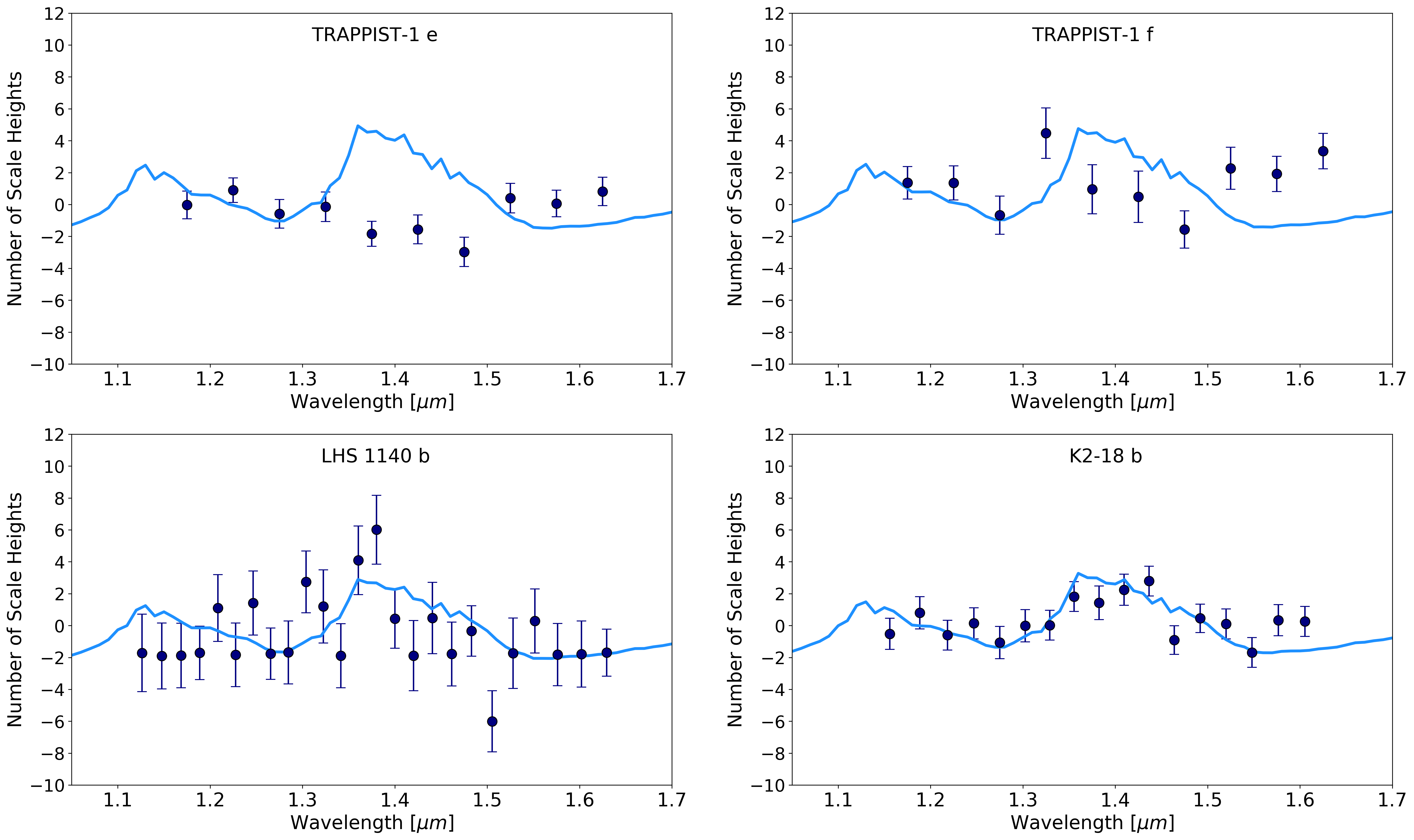}
    \caption{Comparison of the spectrum derived here for LHS\,1140\,b and those of K2-18\,b \citep{Tsiaras_k2-18} as well as TRAPPIST-1 e and f \citep{de_Wit_2018_trappist}. Overplotted for each is a forward model assuming a clear H/He dominated atmosphere with a water abundance of $\log_{10}(V_{\text{H}_{2}\text{O}}) = -3$.}
    \label{fig:trap_comp}
\end{figure*}

The choice of limb darkening coefficients can also have profound effects on the recovered spectrum, particularly for cooler stars \citep[e.g.][]{Kreidberg_GJ1214b_clouds}, and several different limb darkening laws are available. We attempted fits with pre-computed linear, square-root and quadratic coefficients, again calculated using ExoTETHyS \citep{morello_exotethys}, but only the linear values provided usable fits to the white light curve. Additionally, we fitted the data with claret coefficients for different stellar temperatures. The linear law and additional claret fits all resulted in spectra which agree with the one originally derived to within 1$\sigma$, as shown in Figure \ref{fig:extra_ldc}. Nevertheless, we performed retrievals on these spectra, finding they preferred a solution with water to 2.98 and 2.92 $\sigma$ for claret coefficients at 3150 and 3250 K respectively while the linear coefficients resulted in a 3.22 $\sigma$ detection of water.

Finally, a recent study used ESPRESSO and TESS data to provide updated system parameters \citep{lillo_lhs_water}. Hence we tried fitting the light curves using their parameters ($a/R_s$ = 96.4, i = 89.877$^\circ$) as well as performing retrievals using their revised mass (M$_p$ = 6.38 $M_\oplus$). The resultant spectrum, and best-fit retrieval, is shown in Figure \ref{fig:new_para}. The spectrum still shows evidence for water with an abundance of $\log_{10}(\text{V}_{\text{H}_2\text{O}})$ = 3.01$^{+1.59}_{-1.58}$ and a significance of 2.57$\sigma$.

\section{Discussion}

Our results for LHS\,1140\,b prefer the presence of an atmosphere containing water vapour. However, given the noise and scatter of the signal, a flat spectrum cannot be ruled out and the primary/secondary nature of the atmosphere cannot be determined. Additionally, M-dwarfs are known to be capable of creating spectral signatures which can alter the derived atmospheric composition.

Other rocky habitable zone planets have shown a completely flat spectrum \citep{de_Wit_2016_trappist, de_Wit_2018_trappist,ducrot_trap,zhang,wakeford_trap}. The analysis of K2-18\,b \citep{Tsiaras_k2-18, benneke_k2-18} suggests that the planet could have a significant amount of water vapour in the atmosphere but may also have a non-negligible fraction of hydrogen. The presence of water vapour sparked an intense debate on habitability, largely due to the size of the planet and the presence of hydrogen in the atmosphere \citep[e.g.][]{madhu_k218}. A similar detection in the atmosphere of LHS\,1140\,b would have important consequences for habitability, indicating that water vapour is not a rare outcome for smaller, temperate planets. Figure \ref{fig:trap_comp} highlights the HST WFC3 spectra for TRAPPIST-1\,e and f, the planets in system with the most similar equilibrium temperatures to LHS\,1140\,b, as well as K2-18\,b. In each case, a model for a clear primary atmosphere is over-plotted with a water abundance of $\log_{10}(V_{\text{H}_2\text{O}})$ = -3. The TRAPPIST planets clearly do not possess such an atmosphere while the water feature seen on K2-18\,b is far better defined than the potential one uncovered here.

\begin{figure*}
    \centering
    \includegraphics[width=0.46\textwidth]{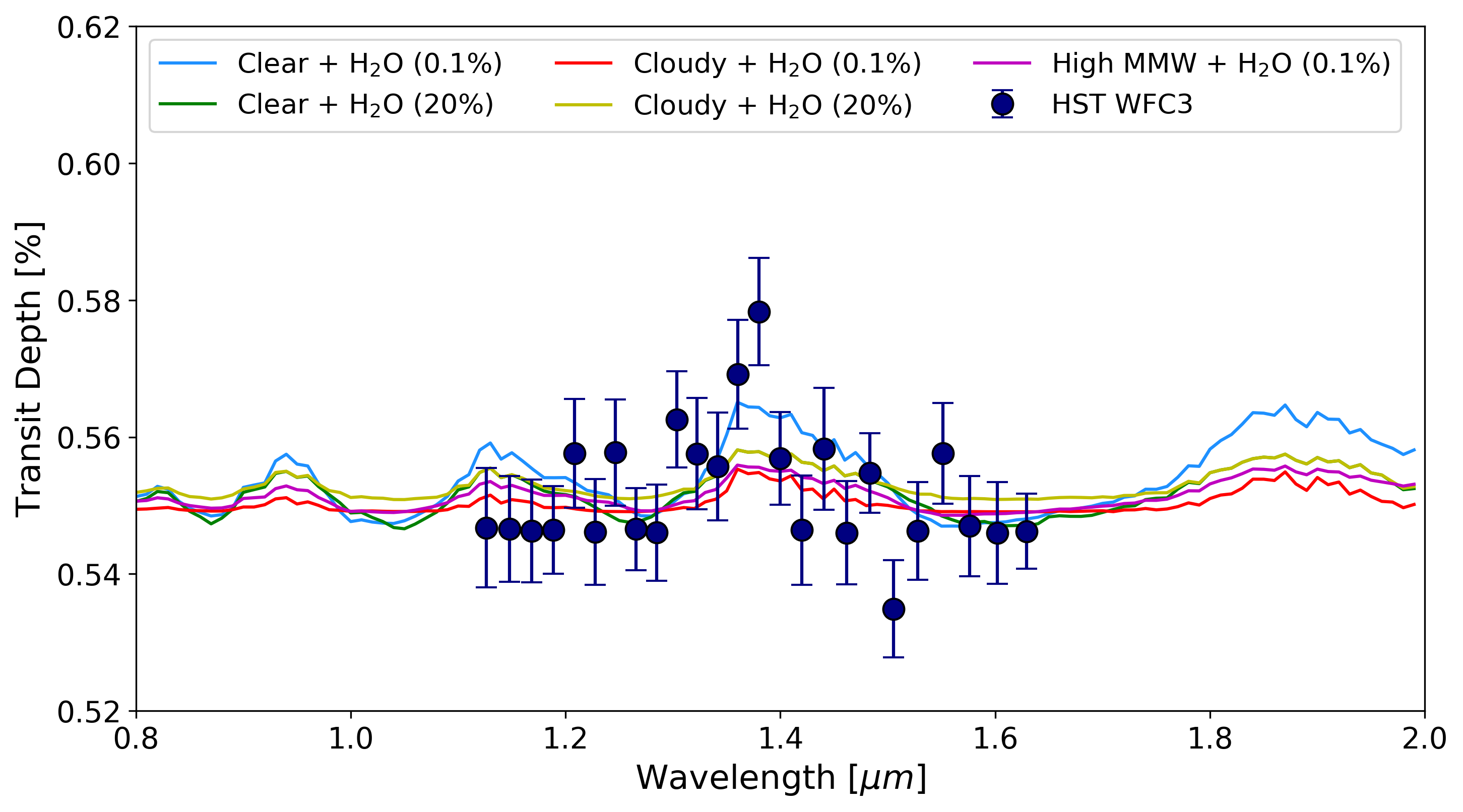}
    \includegraphics[width=0.46\textwidth]{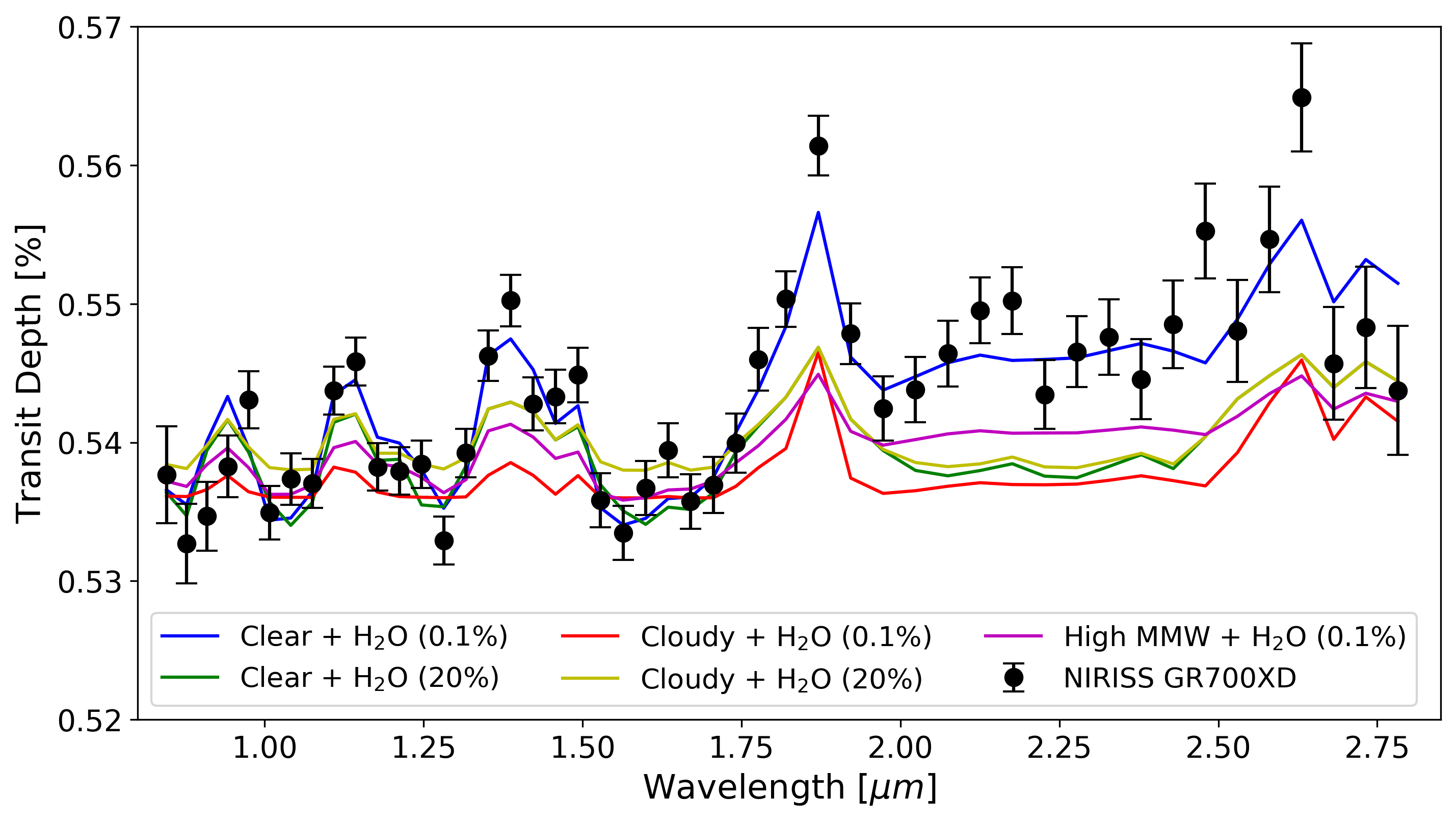}
    \includegraphics[width=0.46\textwidth]{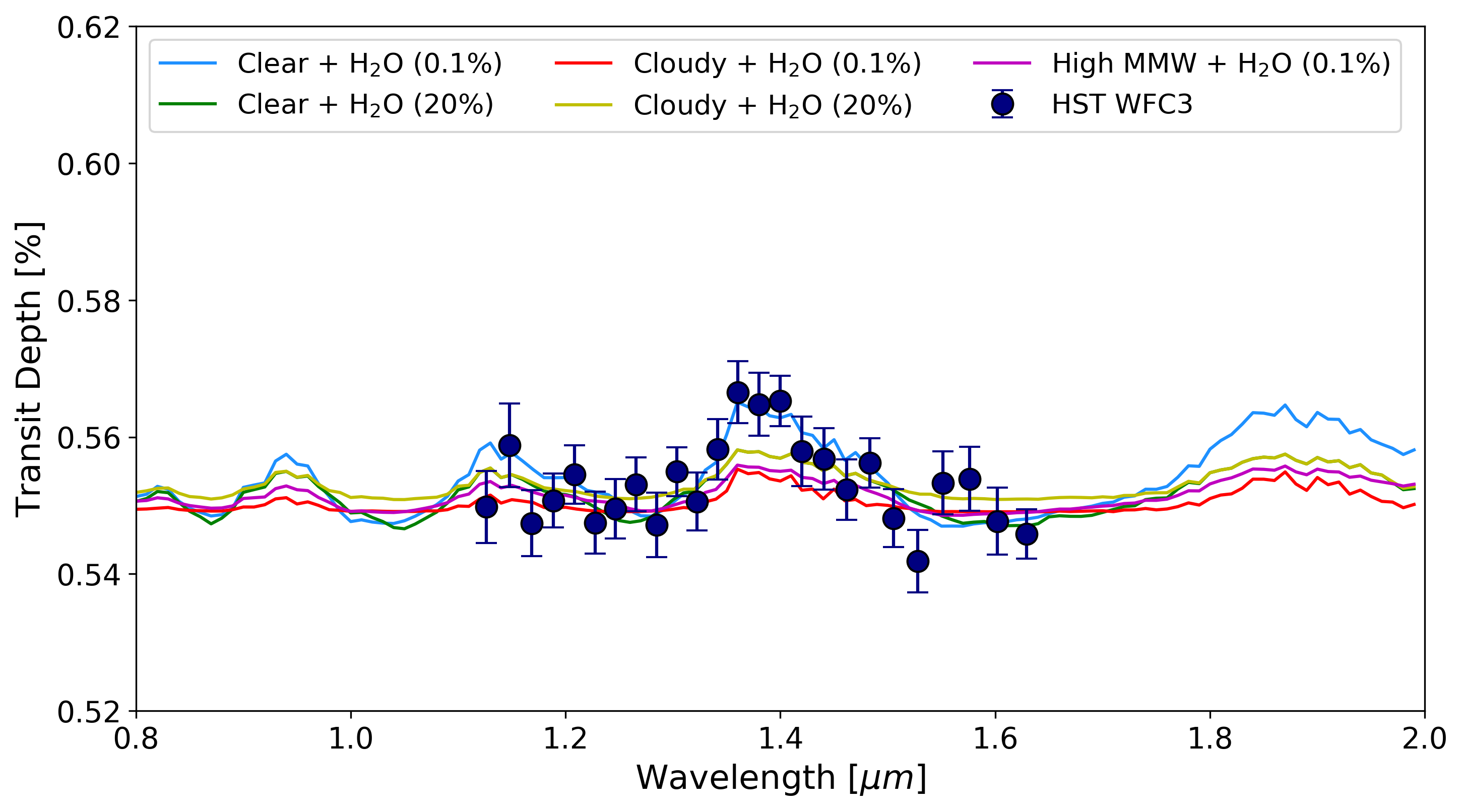}
    \includegraphics[width=0.46\textwidth]{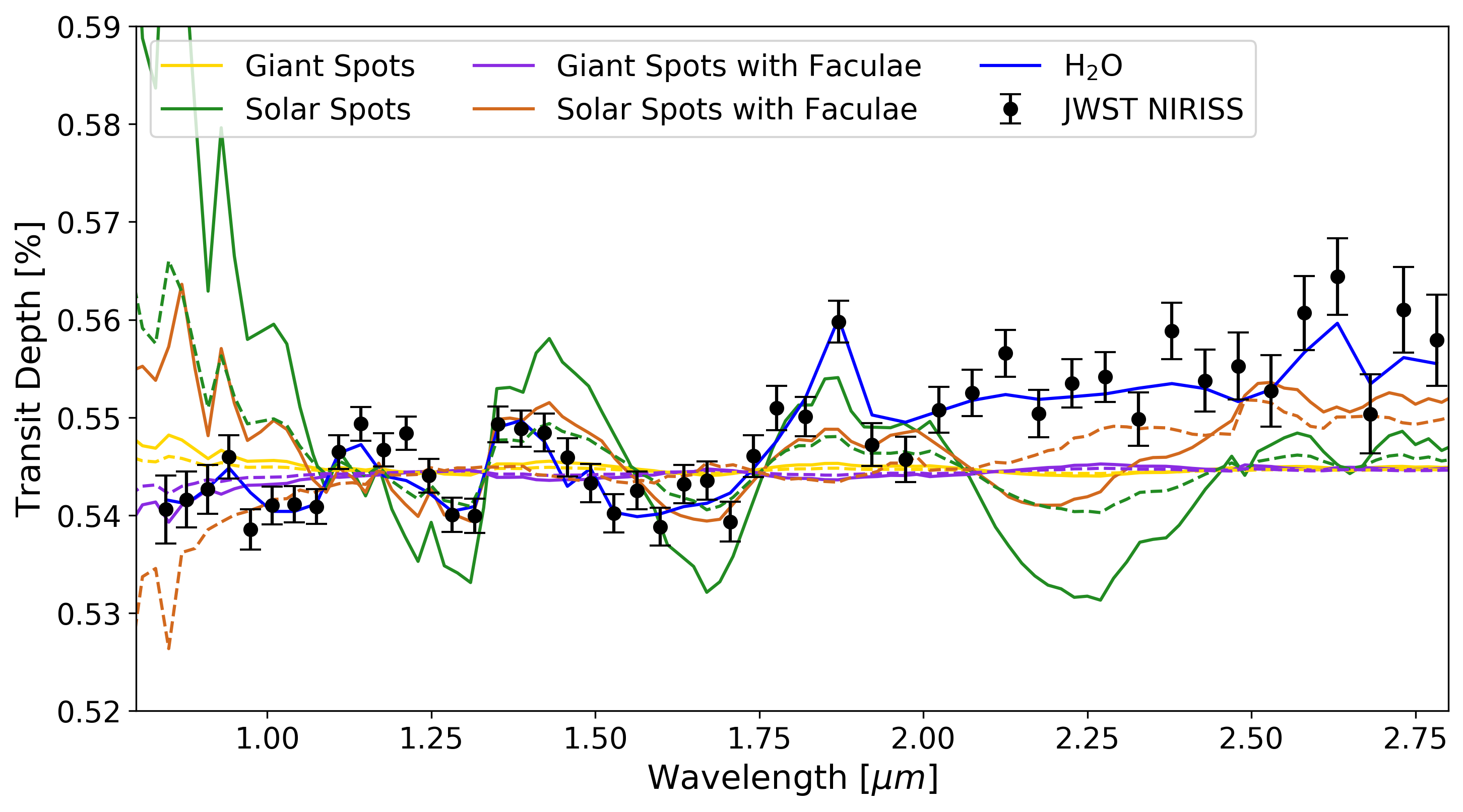}
    \caption{Upper left: Transmission spectrum from the WFC3 G141 data of LHS\,1140\,b analysed here (black) with various forward models over-plotted. The best-fit solution from an atmospheric retrieval favours the presence of a clear, H/He atmosphere with water (blue). Lower left: Simulated final spectrum (black) when the current data is combined with two further Hubble WFC3 G141 observations, increasing the ability to confirm the presence of water. Upper right: Expected data from JWST NIRISS GR700XD which suggests these observations could provide a definitive answer to the question: does LHS\,1140\,b have a clear, H/He dominated atmosphere. Lower right: transit light source models for the JWST data which are heavily intertwined with the atmospheric signal, highlighting the importance of understanding this effect in the JWST era.}
    \label{fig:spectrum}
\end{figure*}

Although it is larger than the TRAPPIST-1 planets, LHS\,1140\,b has a density which is compatible with a rocky composition, predominantly composed of iron and magnesium silicates \citep{dittman_lhs1140b,Ment_mass-radius_2019}. Hence, the presence/absence of a hydrogen envelope around this planet would substantially inform the debate around the TRAPPIST-1 planets and K2-18\,b. The relatively low level and stability of UV flux experienced by LHS\,1140\,b should be favourable for its present-day habitability \citep{spinelli_lhs_uv}, making this planet one of the most interesting targets for the search of bio-signatures in the future although Galactic cosmic rays could impact this \citep{herbst_lhs}. Measurements and interior modelling by \citet{lillo_lhs_water} suggest that the planet could possess a substantial mass of water. Modelling suggests that, if it were to have a surface ocean, LHS\,1140\,b may be in a snowball state \citep{yang_snowball} or have a unglaciated substellar region \cite{checlair_snowball}.

Confidently detecting the presence of spectroscopic signatures will allow us to differentiate between hydrogen rich and heavier atmospheres, a key sign of Super-Earths' provenance and evolution. From a formation perspective, while in-situ formation of Super-Earths is theoretically possible, it may happen only under very specific conditions \citep{Ikoma2012, Ogihara2015}. Current formation models predict that Neptune-mass planets or larger are forced to move to closer orbits when a critical mass is being accreted \citep{Ida2008, Mordasini2009, Ida2010}. Super-Earths could therefore be the remnants of larger planets which have lost part of their initial gaseous envelope, due to XUV-driven hydrogen mass-loss coupled with planetary thermal evolution \citep[e.g.][]{Leitzinger2011, Owen2012, Lopez2012, Owen2013, Owen2017}.

In this scenario planets with radii $< 1.7 \,R_\oplus$ are expected to have lost most of the their primordial hydrogen envelope, while planets larger than $1.7\,R_\oplus$ are expected to have retained at least some of it. Current observations of small, rocky worlds have not been compatible with hydrogen dominated, cloud free atmospheres. With a radius of $1.7\,R_\oplus$, LHS\,1140\,b is situated between these two populations. By being a world which unambiguously has a solid surface, the confirmation/rejection for the presence of large amount of hydrogen would significantly impact our understanding of small worlds. Meanwhile, the potential for large amounts of water in the atmospheres of LHS\,1140\,b and K2-18\,b could hint towards the existence of water worlds \citep{zeng_ww}. Definitive constraints on the atmospheric chemistry of LHS\,1140\,b would inform formation processes, exoplanet evolution and interior/atmospheric models.


\section{Future Observations}

The confirmation/refutation of an atmospheric envelope and of a water signature of LHS\,1140\,b will significantly guide current debates into the nature of small exoplanets, constrain planet evolution models and inform us about the potential habitability of rocky worlds orbiting M-dwarfs. A number of different space-based facilities could be utilised for this study.

Firstly, additional HST WFC3 G141 observations could be taken. Using the errors from data analysis here, we simulated the effect of adding two further Hubble observations based upon the best-fit solution and that these could increase the significance of the detection, in comparison to the flat model, to $>$ 5$\sigma$. Figure \ref{fig:spectrum} displays the spectrum recovered from Hubble WFC3 G141 along with several forward models for a cloudy H/He atmosphere and one with a high mean molecular weight. The addition of two new transits would decrease the average error from $\sim$80 ppm to less than 50 ppm. 

However, disentangling potential stellar contamination will still be difficult given the narrow wavelength coverage. Observations with the G102 grism could help by filling the spectral gap between the current HST data and that from the ground. Nevertheless, given the long baseline between observations, difficulties may still remain. Furthermore, if the atmosphere of LHS\,1140\,b is not clear and hydrogen dominated, then distinguishing between a cloudy primary atmosphere or one with a higher mean molecular weight would be difficult with additional HST data alone.

The James Webb Space Telescope (JWST), currently scheduled for launch in late 2021, will provide unparalleled sensitivity and previous simulations have shown that H$_2$O, CH$_4$ and CO$_2$ could be detected by JWST in the atmosphere of an Earth-like planet around LHS\,1140 \citep{lhs_bio}. We simulate a single transit observation with NIRISS GR700XD and spectrally bin the data to reduce the resolution to R$\sim$50, as shown in Figure \ref{fig:spectrum}. Such a dataset would allow us to confirm, or refute, the presence of a clear, H/He atmosphere around LHS\,1140\,b as it would provide a higher signal to noise ratio on the atmosphere while the wide spectral coverage will probe multiple bands of molecular absorption. The continuous coverage from the visible into the near-infrared would also aid in the fitting of the stellar signal. Constraining an atmosphere with a higher mean molecular weight is likely to take many more transits. Given its long period and location close to the ecliptic plane, only around 4 transits of LHS\,1140\,b per year would be observable with JWST and thus initial observations should be taken as soon as possible. Observations with NIRISS GR700XD were planned as part of GTO Proposal 1201 (PI: David Lafreniere) but were withdrawn in favour of studying other planets \citep{lafreniere_jwst}.

The ESA M4 mission Ariel will survey a population of 1000 exoplanets during its primary mission \citep{tinetti_ariel} and much of this could be dedicated to studying smaller planets in, and around, the radius gap \citep{edwards_ariel}. Although not modelled here, its wide, simultaneous spectral coverage ($0.5-7.8 \mu$m) will undoubtedly be useful for characterising and removing stellar contamination, allowing an accurate recovery of the atmospheric parameters. The same is true for Twinkle, another space-based spectroscopic mission, which will simultaneously cover $0.4-4.5 \mu$m \citep{edwards_twinkle} and both these observatories are likely to be able to observe around $3-4$ transits per year.

Finally, observations monitoring the star, such as those conducted for GJ\,1214  \citep[e.g.][]{berta_gj1214,narita_gj1214,mallonn_gj1214}, are also important to place further constrains on the spot covering fraction and thus to estimate, and correct for, the transit light source effect \citep[e.g.][]{rosich_correction}. Such campaigns will be vital for all upcoming observations of smaller, temperature worlds, particularly for those around cooler stars \citep{apai_stellar_cont}. 

Long-term monitoring of LHS\,1140 is especially crucial given the system also hosts a smaller (1.28 $R_\oplus$) but warmer (440 K) terrestrial planet \citep{Ment_mass-radius_2019,feng_lhs} which, due to it's higher equilibrium temperature and lower surface gravity, could have larger spectral features and studies with future missions could allow comparative planetology within the same system.

\section{Conclusions}

Observing rocky, habitable-zone planets pushes the limits of our current technology. We have presented the HST WFC3 transmission spectrum of such a world and shown that it could be compatible with the presence of an atmosphere containing water. A popular aphorism suggests the magnitude of a claim should be balanced by the weight of the evidence. Here, however, the evidence for water is marginal compared to a flat model, the primary or secondary nature of the atmosphere cannot be determined, and the signal could be distorted by stellar contamination. Hence more data is required to substantiate this claim and unveil the true nature of the planet, with future observatories possessing the power to do so.

\vspace{3mm}
\textbf{Software:} Iraclis \citep{tsiaras_hd209}, TauREx3 \citep{al-refaie_taurex3}, pylightcurve \citep{tsiaras_plc}, ExoTETHyS \citep{morello_exotethys}, Astropy \citep{astropy}, h5py \citep{hdf5_collette}, emcee \citep{emcee}, Matplotlib \citep{Hunter_matplotlib}, Multinest \citep{Feroz_multinest}, Pandas \citep{mckinney_pandas}, Numpy \citep{oliphant_numpy}, SciPy \citep{scipy}, corner \citep{corner}.

\vspace{3mm}
\textbf{Data:} This work is based upon observations with the NASA/ESA Hubble Space Telescope, obtained at the Space Telescope Science Institute (STScI) operated by the Association of Universities for Research in Astronomy, Inc. under NASA contract NAS 5-26555. The publicly available HST observations presented here were taken as part of proposal 14888, led by Jason Dittmann. These were obtained from the Hubble Archive which is part of the Mikulski Archive for Space Telescopes. This paper includes data collected by the TESS mission which is funded by the NASA Explorer Program. TESS data is also publicly available via the Mikulski Archive for Space Telescopes (MAST).

\vspace{3mm}
\textbf{Acknowledgements:} We thank our referee for insightful comments and constructive discussions which improved the quality of the manuscript. This project has received funding from the European Research Council (ERC) under the European Union's Horizon 2020 research and innovation programme (grant agreement No 758892, ExoAI) and under the European Union's Seventh Framework Programme (FP7/2007-2013)/ ERC grant agreement numbers 617119 (ExoLights). Furthermore, we acknowledge funding by the Science and Technology Funding Council (STFC) grants: ST/K502406/1, ST/P000282/1, ST/P002153/1 and ST/S002634/1. Finally, this work was supported by Grant-in-Aid for JSPS Fellows, Grant Number JP20J21872.

\bibliographystyle{aasjournal}
\bibliography{main}

\end{document}